\documentclass[sigconf]{acmart}
\AtBeginDocument{%
  }

\copyrightyear{2026}
\acmYear{2026}
\setcopyright{cc}
\setcctype{by}
\acmConference[CHI '26]{Proceedings of the 2026 CHI Conference on Human Factors in Computing Systems}{April 13--17, 2026}{Barcelona, Spain}
\acmBooktitle{Proceedings of the 2026 CHI Conference on Human Factors in Computing Systems (CHI '26), April 13--17, 2026, Barcelona, Spain}
\acmPrice{}
\acmDOI{10.1145/3772318.3790667}
\acmISBN{979-8-4007-2278-3/2026/04}

\usepackage{float}
\usepackage{dirtytalk}
\usepackage{subcaption}
\usepackage[normalem]{ulem}
\begin{document}
\newcommand{\new}[1]{\textcolor{black}{#1}}

\title[AttentiveLearn]{AttentiveLearn: Personalized Post-Lecture Support for Gaze-Aware Immersive Learning}

\author{Shi Liu}
\affiliation{%
  \institution{Karlsruhe Institute of Technology}
  \city{Karlsruhe}
  \country{Germany}}
\email{shi.liu@kit.edu}

\author{Martin Feick}
\affiliation{%
  \institution{Karlsruhe Institute of Technology}
  \city{Karlsruhe}
  \country{Germany}}
\email{martin.feick@kit.edu}

\author{Linus Bierhoff}
\affiliation{%
  \institution{Karlsruhe Institute of Technology}
  \city{Karlsruhe}
  \country{Germany}}
\email{linus.bierhoffk@student.kit.edu}

\author{Alexander Maedche}
\affiliation{%
  \institution{Karlsruhe Institute of Technology}
  \city{Karlsruhe}
  \country{Germany}}
\email{alexander.maedche@kit.edu}

\renewcommand{\shortauthors}{Anonymous Authors}

\begin{abstract}
Immersive learning environments such as virtual classrooms in Virtual Reality (VR) offer learners unique learning experiences, yet providing effective learner support remains a challenge. While prior HCI research has explored in-lecture support for immersive learning, little research has been conducted to provide post-lecture support, despite being critical for sustained motivation, engagement, and learning outcomes. To address this, we present \textit{AttentiveLearn}, a learning ecosystem that generates personalized quizzes on a mobile learning assistant based on learners’ attention distribution inferred using eye-tracking in VR lectures. We evaluated the system in a four-week field study with 36 university students attending lectures on Bayesian data analysis. \new{\textit{AttentiveLearn} improved learners’ reported motivation and engagement, without conclusive evidence of learning gains. Meanwhile, anecdotal evidence suggested improvements in attention for certain participants over time.} Based on our findings of the field study, we provide empirical insights and design implications for personalized post-lecture support for immersive learning systems.
\end{abstract}


\begin{teaserfigure}
  \centering
  \includegraphics[width=\textwidth]{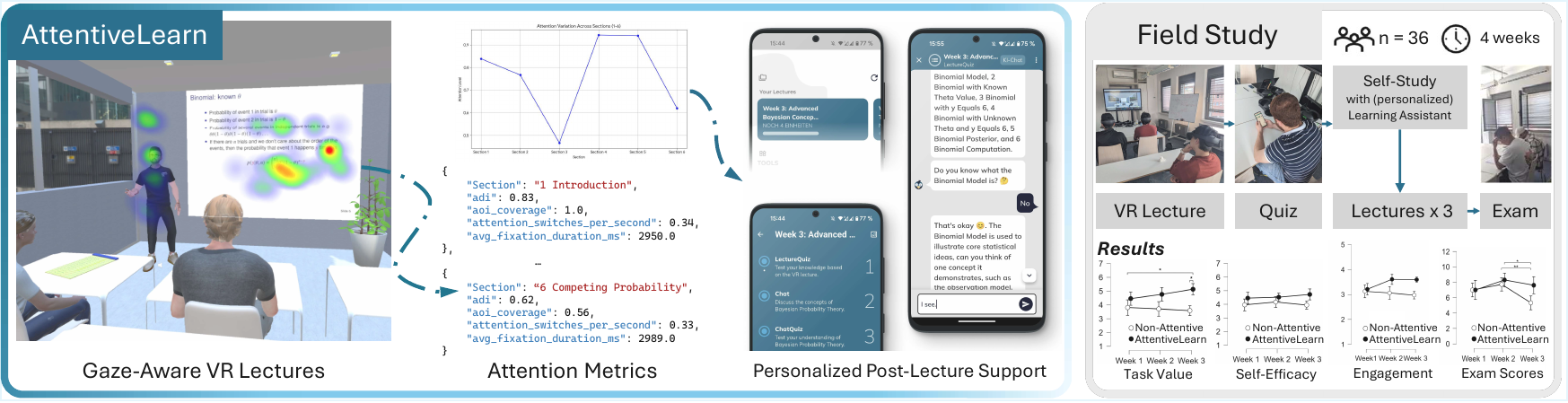}
  \caption{Overview of the \textit{AttentiveLearn} system and study design. The system combines (a) gaze-aware VR lectures, (b) attention metrics, and (c) personalized post-lecture support via an LLM-based mobile assistant. The field study with 36 students over four weeks evaluated the system's impact on engagement, motivation, and learning outcomes in a real-world lecture setting.}
  \Description{Figure shows an overview of the AttentiveLearn system and field study. From left to right, the system pipeline includes: (a) a VR classroom with avatars and a heatmap overlay showing gaze tracking; (b) graphs and JSON-like outputs of computed attention metrics such as AOI coverage and fixation duration; (c) a mobile interface providing personalized post-lecture quiz questions and feedback. On the right, the field study design is depicted: across four weeks, 36 students participated, including VR lectures, quizzes, three lecture sessions, and a final exam. Results are summarized with line plots indicating group comparisons on task value, self-efficacy, engagement, and exam scores.}
  \label{fig:teaser}
\end{teaserfigure}

\maketitle

\section{Introduction}

Using immersive technologies such as Virtual Reality (VR) to create engaging learning experiences is a broad research field known as immersive learning~\cite{mystakidis_immersive_2023, makransky_cognitive_2021}. Prior HCI research has investigated the design of immersive learning systems across various domains, ranging from digital fabrication~\cite{radu_virtual_2021} to language learning~\cite{shao_teaching_2020}. Different affordances and formats of immersive learning have been explored, including virtual laboratories~\cite{radu_virtual_2021}, simulations~\cite{zhu_designing_2025}, and virtual classrooms~\cite{gao_digital_2021}. While immersive learning environments have shown promise for better learning outcomes~\cite{makransky_benefits_2022, baceviciute_investigating_2020}, providing effective learner support in these environments remains a challenge~\cite{shen_educator_2025, petersen_pedagogical_2021}. A key setting where this challenge has been recognized is the immersive virtual classroom in VR~\cite{han_exploring_2022, gao_digital_2021}. Immersive virtual classrooms have emerged as a prominent example of immersive learning, both widely researched in HCI~\cite{gao_digital_2021, liu_classmeta_2024, liu_gazeclass_2025} and increasingly adopted in practice~\cite{moynihan_immersive_2021}. Prior work has examined design features such as embodied agents~\cite{liu_classmeta_2024}, large language model (LLM) integration~\cite{tang_llm_2025}, and annotation tools~\cite{tsai_gazenoter_2025} to optimize the learning experience. However, most of these efforts focus on in-lecture support, whereas post-lecture support for immersive virtual classrooms has received less attention.

Post-lecture support is well established as a critical component of learning success~\cite{palmer_comparison_2019}. Educational research emphasizes that learning is a continuous and ubiquitous process extending beyond the classroom~\cite{kuh_guiding_1996,hwang_research_2011}. Techniques such as tutoring~\cite{dmello_autotutor_2012}, quizzes~\cite{raes_learning_2020}, \new{content reviews~\cite{szafir_artful_2013}, }and other follow-up activities have been applied to improve learning outcomes. They have been shown to be particularly effective when tailored to individual learners' prior experiences and behaviors in lectures~\cite{du_plooy_personalized_2024, ross_adaptive_2018, szafir_artful_2013}. \new{For example, biosignal-based attention metrics have been used to drive adaptive content reviews for desktop-based learning systems~\cite{szafir_artful_2013}. We extend this approach to immersive virtual classrooms.} In immersive learning, however, much of the individual learning behavior, preferences, and performance exhibited during the lecture is often lost, making it difficult to provide meaningful personalized support afterward. We identify this as a research gap: while personalized post-lecture support based on in-lecture metrics has shown potential to enhance motivation~\cite{ross_adaptive_2018}, engagement~\cite{nuci_game-based_2021}, and learning outcomes~\cite{ross_adaptive_2018}, its role in immersive learning contexts remains underexplored\new{, lacking a framework to guide the transfer of in-situ attention monitoring to ex-situ support.}

\new{To address this gap, we ground our framework and system design in two key streams of prior research}. First, research has revealed that focused attention during lectures is a crucial cognitive factor for learning success~\cite{keller_attention_2020}. In HCI research, eye-tracking technologies have been widely used to support attention across diverse contexts~\cite{duchowski_eye_2007, roda_attention_2006}, including education~\cite{dmello_gaze_2012, hutt_out_2017} and in VR systems~\cite{plopski_eye_2022, han_exploring_2022}. Yet, little research has explored how attention data can inform post-lecture support in immersive learning. Second, we identify quizzes as a pedagogically effective post-lecture support strategy. Quizzes are widely used to reinforce learning, provide self-assessment, and improve retention through the testing effect~\cite{roediger_iii_test-enhanced_2006}. They have also been shown to increase engagement~\cite{raes_learning_2020}. Moreover, adaptive and personalized quizzes have been explored as promising post-lecture scaffolds~\cite{ross_adaptive_2018}. However, prior work has not investigated using attention data from VR lectures as a personalization strategy for post-lecture quizzes.

Building on these rationales, we introduce \new{a framework for attention-based post-lecture personalization in the context of immersive learning, instantiated with} \textit{AttentiveLearn}, a learning ecosystem that generates personalized quiz questions based on learners' attention distribution during VR lectures. The ecosystem integrates VR lectures with an existing mobile learning assistant application deployed at our local university. \textit{AttentiveLearn} \new{operationalizes the framework through} three components: (a) a gaze-aware VR classroom that collects eye-tracking data during lectures, (b) a data processing pipeline that computes attention metrics and transfers them to an external server, and (c) a mobile learning assistant application that delivers post-lecture support. 

To evaluate \textit{AttentiveLearn}, we conducted a four-week between-subjects field study with 36 students. Each week, participants attended an on-site VR lecture on Bayesian data analysis. After the lecture, half of the participants used a variant of the mobile assistant with quizzes personalized using eye-tracking data collected during the lecture, while the other half used a version with randomly generated, non-attentive quiz questions. We collected survey data on motivation and engagement each week, as well as results from intermediate and final exams to evaluate learning outcomes. In addition, we analyzed interaction and chat logs of the learning assistant and conducted 10 semi-structured interviews to provide qualitative insights. Our findings suggest that the personalized assistant improved reported engagement and motivation throughout the study. There was also anecdotal evidence of improved attention for low-attention learners from the first to the last week. Learning outcomes showed improvement in the intermediate tests, though no significant differences were observed in the final test. In this work, we make the following three key contributions: 

\begin{itemize}
\item Conceptualization of a novel learning ecosystem \new{with a generalizable framework} that bridges in-situ immersive learning with ex-situ support through attention-aware post-lecture quizzes.
\item Integration of the attention-aware personalization pipeline into an existing mobile learning assistant, demonstrating feasibility in real-world educational settings.
\item Empirical insights and design implications from a four-week field study ($n = 36$) investigating how personalized post-lecture support affects motivation, engagement, and learning outcomes.  
\end{itemize}

\section{Related Work}
\label{sec:RW}
\subsection{Immersive Learning Environments}
Immersive learning refers to the use of technologies, such as VR and Augmented Reality (AR), to create engaging, situated learning experiences that support embodied cognition and enhance learning outcomes~\cite{mystakidis_immersive_2023, makransky_cognitive_2021}. Empirical evaluations of immersive learning systems reveal a broad range of learning benefits. For instance, a VR field study conducted by \citet{petersen_pedagogical_2021} demonstrated improved enjoyment and knowledge acquisition in the context of museum guides. Moreover, \citet{radu_what_2019} identified students' increased self-efficacy after using an immersive physics learning application. Other positive impacts include improved comprehension~\cite{schnitzer_language_2025}, motivation~\cite{thanyadit_tutor_2023}, and achievement across contexts~\cite{radianti_systematic_2020}. However, research also reports mixed or contradictory effects~\cite{makransky_immersive_2021, radianti_systematic_2020}, which indicate unaddressed user challenges in immersive learning environments. Therefore, recent systematic reviews of immersive learning systems highlight the importance of cognitive \new{and learner-centered} support as a critical determinant of learning and motivation~\cite{makransky_cognitive_2021}. \new{To address the need for learner-centered support, \citet{palmas_acceptance_2019} investigated public‑speaking training in VR and how skill development can be supported through direct feedback and presence-enhancing mechanisms~\cite{palmas_virtual_2021}.}

As one form of immersive learning environments, virtual classrooms implemented in VR have attracted increasing attention in HCI. Researchers have leveraged VR-specific affordances, such as embodied agents~\cite{liu_classmeta_2024} and VR videos as virtual excursions~\cite{cheng_case_2019}, to enhance engagement and interactivity. To further enrich the classroom experience, prior work has introduced both teacher-centric tools (e.g., authoring tools~\cite{shen_educator_2025}, teaching augmentation~\cite{an_ta_2020}) and learner-centered support mechanisms~\cite{liu_classmeta_2024}. Our work aligns with the latter, focusing on learner-centered support in virtual classrooms. \new{While some of these support techniques have analogs in desktop or video-based learning~\cite{szafir_artful_2013, thanyadit_tutor_2023}, VR introduces unique immersive dynamics, such as spatial attention shifts and embodied presence, making dedicated support for immersive learning necessary~\cite{liu_classmeta_2024}.}

\new{For immersive virtual classrooms, }learner-centered designs have included embodied agents acting as virtual peers during lectures~\cite{liu_classmeta_2024}, as well as studies of classroom dynamics such as avatar representation, teacher-student proximity, and classroom layout~\cite{gao_digital_2021, blume_students_2019}. However, the majority of this work has emphasized in-lecture support, with relatively little attention to how support can extend beyond the immersive lecture itself. Among the in-lecture support studies, a few have explored adaptive and cognitive support, for example, through gaze-based tools for note-taking~\cite{tsai_gazenoter_2025} and attention-based warnings during lectures~\cite{han_exploring_2022}, pointing toward promising directions for post-lecture support in immersive learning environments. \new{Targeting the virtual classroom as our design space, our work takes a first step toward transferring established attention-support techniques into immersive learning, while acknowledging the need for investigating learner-centered support in more embodied and immersive learning scenarios in future research.}

\textbf{In summary, immersive virtual classrooms have shown promise, but learner-centered support often stops at in-lecture support, leaving open the question of how learning can be effectively extended and supported after the session.}

\subsection{Attention Aware Systems and Gaze-adaptive Support}
To provide cognitive support in learning contexts, attention-aware systems focus on one key cognitive process of attention~\cite{roda_attention_2006}. Attention has been defined as the process of selecting relevant perceived information, allowing individuals to become ``active seekers and processors'' of knowledge~\cite{chun_visual_2008}. In learning, attention is essential for effective information processing~\cite{mayer_cognitive_2014} and sustaining situational interest~\cite{makransky_cognitive_2021}. In the cognitive-affective model of immersive learning proposed by \citet{makransky_cognitive_2021}, attention is identified as a crucial factor influencing learning outcomes.

HCI research has developed attention-aware systems to scaffold attentional processes~\cite{roda_attention_2006}. A closely related concept is attentive user interfaces, which focus on making the interface responsive to users’ attentional states, rather than treating attention support as the core design goal of the system~\cite{vertegaal_designing_2006, bulling_pervasive_2016}. Regardless of the nuanced differences between the two concepts, eye-tracking is the most widely used mechanism for these systems, offering a non-intrusive proxy for user attention~\cite{duchowski_eye_2007}. In existing research, attention-aware systems have included gaze-adaptive intelligent tutoring~\cite{dmello_gaze_2012}, note-taking~\cite{khan_gaze_2019}, reading \& writing support~\cite{10.1145/2070719.2070722,langner_leveraging_2023}, and adaptive feedback during learning tasks~\cite{jarodzka_tracking_2017, liu_af-mix_2025}. \new{Besides these gaze-adaptive systems, EEG-based cognitive support and attention monitoring have also been explored~\cite{zander_enhancing_2010, roda_attention_2006}.}

Specifically for lecture consumption and in classroom settings, \citet{hutt_automated_2019} used eye-tracking to detect mind wandering during recorded lectures, demonstrating the feasibility of in-lecture attention monitoring. \citet{10.1145/3706598.3713480} developed a learning feedback system that aggregates the gaze data of peers to inform the area of interests (AOI) during lectures. \new{Furthermore, \citet{szafir_artful_2013} developed \textit{ARTFul}, a system that leveraged EEG to monitor attention during lectures and provide learners with suggestions for content review. Their results demonstrated that with personalized content review, learners achieve better learning and recall. Their work established that attention signals can effectively guide post-lecture support and identified several future research directions which we address. First, they acknowledged the limitations of passive content review and advocated for exploring \say{other embedded assessment techniques} based on attention. We respond to this by using personalized quizzes administered both directly after the lecture and before the subsequent lecture, aiming to provide more comprehensive and active support. Second, they envisioned that advances in intelligent content summarization could provide learners with a \say{truly customized educational experience}; we design our system towards this by leveraging LLMs for quiz generation and post-lecture Q\&A chat. Lastly, they noted the limitations of EEG, including that it's prone to be affected by extraneous signals. Within our target design space of immersive learning support, we employ eye-tracking, which is increasingly integrated into VR headsets and seen as a less obtrusive sensing method~\cite{plopski_eye_2022}.} Eye-tracking has been established as a reliable and non-intrusive approach for attention-aware support~\cite{chun_visual_2008, duchowski_eye_2007, roda_attention_2006}. More broadly, researchers have begun to examine the potential of using gaze data for personalized content generation when combined with LLMs. For instance, \citet{abdrabou_gaze_2025} discussed both opportunities and ethical concerns of gaze-informed LLM systems, though their work was not situated in VR or immersive learning. \new{Therefore, compared to existing attention-based support for non-immersive learning using EEG~\cite{szafir_artful_2013}, the open question we address is how to (a) capture attention more unobtrusively in immersive environments, (b) transfer those signals across devices after the lecture session, and (c) translate attention into personalized quizzes rather than content replays.}

In immersive learning environments, however, leveraging gaze data for cognitive support has only recently emerged as a research focus. Recent explorations include the framework by \citet{abeysinghe_framework_2025}, who proposed measuring attention with eye-tracking in immersive settings to adapt the presentation of learning materials. While promising, their work remained limited to technical feasibility and did not address learner-centered content adaptation or post-lecture support. Similarly, \citet{han_exploring_2022} and \citet{tsai_gazenoter_2025} designed gaze-adaptive support mechanisms within VR sessions, such as adaptive hints and note-taking aids. Furthermore, \citet{liu_classmeta_2024} integrated LLM-driven embodied agents to support students during VR lectures. However, none of these works have addressed gaze-informed personalization beyond the session itself.

\textbf{Thus, while gaze-adaptive support in attention-aware systems demonstrates strong potential, prior work largely focuses on in-situ adaptation, leaving the post-lecture support underexplored, especially in immersive learning contexts.}

\subsection{Personalized Quizzes and Learning Assistants}
Personalized support after lectures has long been explored in educational research, with quizzes recognized as one validated strategy for learning reinforcement. Quizzes have been shown to enhance metacognition and self-regulation~\cite{dunlosky_improving_2013}. Besides, the testing effect of post-lecture quizzes has been widely observed to improve knowledge retention~\cite{mcdaniel_test-enhanced_2011}, self-efficacy~\cite{dunlosky_improving_2013}, and engagement~\cite{nuci_game-based_2021}.

Beyond static, pre-defined questions, adaptive and personalized quizzes have also been supported in digital learning platforms. For instance, \citet{ross_adaptive_2018} designed adaptive quizzes that adjusted question difficulty based on prior responses, leading to improvements in motivation and engagement. Similarly, \citet{klaveren_effect_2017} showed that adaptive question sequencing better supported diverse learning needs. More recent work by \citet{contrino_using_2024} integrated personalized quizzes into a smart learning platform that dynamically adjusted course structures. In HCI, adaptive learning and personalization have also been studied in intelligent tutoring systems~\cite{dmello_autotutor_2012}, assessment tools~\cite{gamage_optimising_2019}, personalized learning analytics~\cite{demmans_epp_editorial_2023}, etc. However, personalization in these works typically relies on performance data, with fewer systems incorporating cognitive states such as attention. Moreover, little research has adapted these strategies specifically to virtual classrooms.

In parallel, learning assistants as mobile applications have been widely studied as flexible tools for delivering post-lecture support~\cite{hwang_research_2011}. They are capable of delivering a ubiquitous learning experience independent of human tutors after the lectures~\cite{hwang_research_2011}. Studies have integrated quizzes into mobile assistants~\cite{nuci_game-based_2021}. With AI-driven personalization, recent work has explored LLM-based assistants with adaptive learning support~\cite{wambsganss_arguetutor_2021}. For immersive learning, cross-device interaction research in HCI already underscores the opportunity of extending immersive learning through mobile devices~\cite{brudy_cross-device_2019, zhu_phoneinvr_2024}. Yet, immersive learning research has not sufficiently investigated mobile assistants that extend VR lectures with ex-situ personalized support.

\textbf{In short, while personalized quizzes and mobile learning assistants have been researched, their integration into immersive learning with attention-aware personalization has yet to be explored.}

\section{AttentiveLearn: Designing an Attention-Aware Learning Ecosystem}
\label{sec:design}
After identifying the research gap, we clarified our design goal and scope of \textit{AttentiveLearn}: we aim to provide personalized post-lecture support that adapts to individual attention levels during immersive learning sessions. \new{Rather than innovating in VR classroom design itself, our design serves as a reference implementation of a generalizable framework, focusing on bridging in-situ immersive learning with ex-situ personalized support through a mobile assistant.} Specifically, we investigate how attention-aware personalization influences students’ motivation, engagement, and learning outcomes.

\subsection{An Immersive Virtual Classroom}
\begin{figure}[t]
    \centering
    \includegraphics[width=\linewidth]{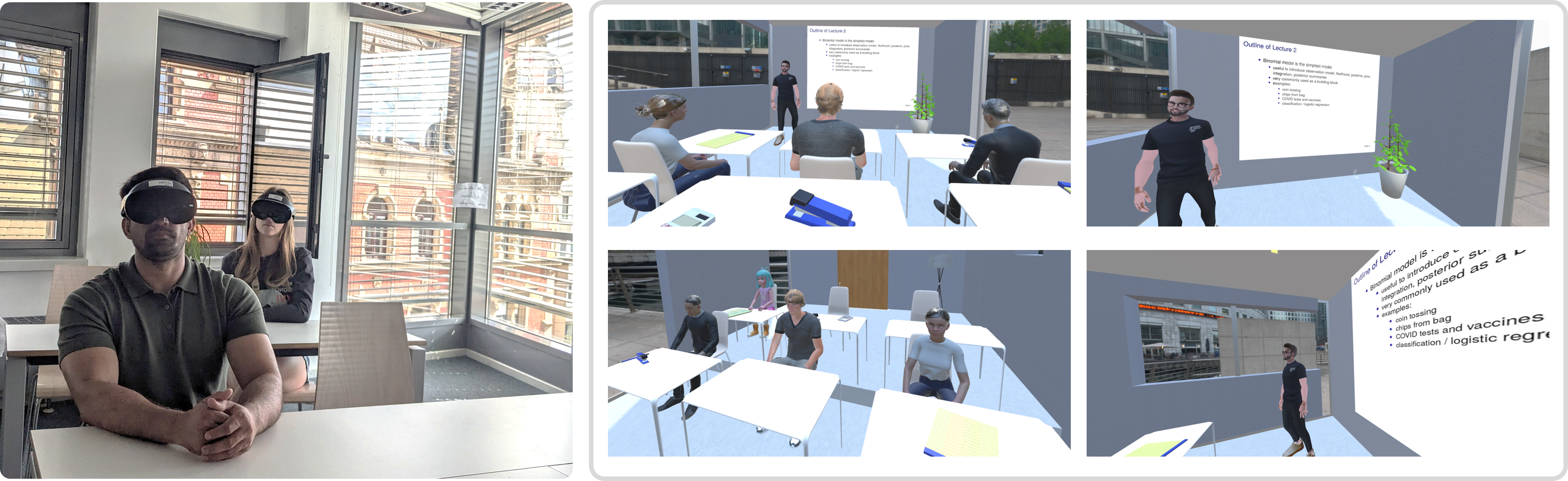}
    \caption{An immersive virtual classroom setup. Left: students using VR headsets in a real-world environment. Right: corresponding classroom scene showing multiple perspectives of the lecture space with avatars and slides.}
    \Description{The figure illustrates the immersive classroom environment. On the left, two students are seated in a physical classroom, wearing VR headsets. On the right, four panels show the corresponding VR classroom scene, adapted from prior work. The VR space includes avatars of students and instructors, a projected lecture slide, desks, and classroom objects such as plants and windows. Different camera angles depict the layout and lecture interaction.}
    \label{fig:VR}
\end{figure}
\new{In our framework, the virtual classroom serves as the \textit{In-Situ Data Acquisition} layer.} Immersive virtual classrooms represent an established problem space rather than a novel contribution of this work. We chose this setting for two reasons: (1) gaze-based attention metrics have been validated in classroom contexts~\cite{hutt_automated_2019, han_exploring_2022}, and (2) virtual classrooms remain a widely adopted format in both HCI research and practice~\cite{moynihan_immersive_2021, gao_digital_2021}, with open-source infrastructures further provide a practical foundation for the design space~\cite{oehlberg_vr_2018}.

For our implementation, we adopted the existing Unity-based virtual classroom from \citet{liu_gazeclass_2025} with their consent. This system provides a standard classroom setup consisting of a slideshow and multiple embodied avatars. In their classroom setup, one avatar represents the lecturer, delivering a pre-recorded lecture with synchronized audio, lip movements, and slide transitions. Additional avatars simulate peer students, creating the impression of a live lecture. Furthermore, the system already integrates a gaze data collection pipeline and a basic gaze-duration–based attention metric (AOI coverage), which has been technically evaluated and aligns with established attention metrics for in-lecture interventions~\cite{han_exploring_2022} and teacher-oriented attention monitoring~\cite{thanyadit_tutor_2023}. This provides a solid foundation for our work. Our modifications to this environment were minimal. We extended the system’s attention tracking with the ability to define AOIs on each slide. This enables tracking of attention switches between AOIs besides the gaze duration. Based on the gaze duration and the number of attention switches during each lecture section (the sections can be predefined in the Unity application via timestamps), an Attention Distribution Index (ADI) can be later calculated for each section in the attention-aware personalization pipeline following \citet{sharma_eye-tracking_2020}.

Thus, the immersive virtual classroom functions as a validated problem space that allows us to investigate our main contribution: \new{designing a framework for attention-aware personalization for post-lecture support.}

\subsection{Attention-Aware Personalization Pipeline}
\begin{figure*}[h]
    \centering
    \includegraphics[width=\linewidth]{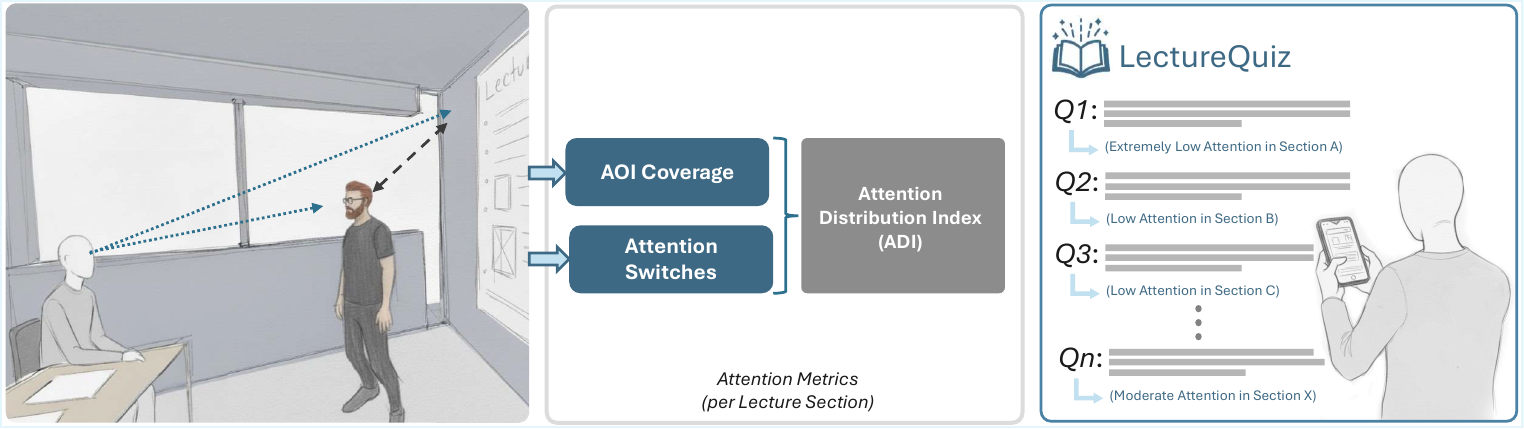}
    \caption{Attention-aware personalization pipeline. Eye-tracking data from VR lectures are processed into attention metrics, which are then used to generate personalized \textit{LectureQuiz} questions targeting low-attention sections.}
    \Description{Figure shows the attention-aware personalization pipeline. On the left, a student in a VR lecture is tracked with gaze rays. In the middle, a processing block computes three attention metrics: AOI coverage, attention switches, and the Attention Distribution Index (ADI). On the right, a student holding a phone receives LectureQuiz questions. Each question is tailored to sections of the lecture where the student’s attention was low or moderate.}
    \label{fig:Pipeline}
\end{figure*}

\paragraph{Data Processing} We implemented a Python-based web server using Flask-SocketIO\footnote{\url{https://flask-socketio.readthedocs.io}} that connects the VR classroom with a mobile assistant application. \new{This pipeline instantiates the framework's \textit{Data Translation} layer, converting raw signals into pedagogical insights. It executes three steps:} (1) receiving raw eye-tracking data with timestamped gaze positions as CSV files from the VR application; (2) processing these into section-level attention metrics; and (3) forwarding metrics in a JSON file to the mobile assistant. The metrics included (a) AOI coverage percentage, (b) number of attention switches, and (c) the ADI metric. AOI coverage was calculated as in \citet{liu_gazeclass_2025}: the accumulated gaze duration on predefined learning-related objects (lecturer avatar and slides) divided by the total lecture length.

Attention switches were approximated from gaze shifts~\cite{sharma_eye-tracking_2020}. The ADI metric combines both measures to represent overall attention across lecture sections, following the methodology of \citet{sharma_eye-tracking_2020} and \citet{hutt_automated_2019}. We intentionally relied on these established measures rather than proposing new ones, \new{to ensure that the pipeline remains adaptable and modular, allowing the framework to accommodate alternative biosignals or attention definitions in future iterations.}

\paragraph{LectureQuiz} These metrics inform an attention-aware quiz module named \textit{LectureQuiz}, which provides immediate feedback and assessment after each lecture, following the principle of test-enhanced learning~\cite{mcdaniel_test-enhanced_2011}. Personalization works as follows: \textit{LectureQuiz} provides learners with LLM-generated questions focusing on sections with low AOI coverage and ADI (see Supplementary Materials for prompts). The number and difficulty of questions can be configured in the prompts based on the learning objectives. In our study, we investigate the integration of \textit{LectureQuiz} in a mobile application, while also validating the perceived quality of \textit{LectureQuiz} (see Section \ref{sec:quizresult}). This integration ensures that attention-aware personalization is not an isolated design feature but part of a comprehensive learning ecosystem.

\subsection{Integrating Personalized Quizzes in a Mobile Assistant}
\label{sec:amsldesign}
\begin{figure*}[h]
    \centering
    \includegraphics[width=\linewidth]{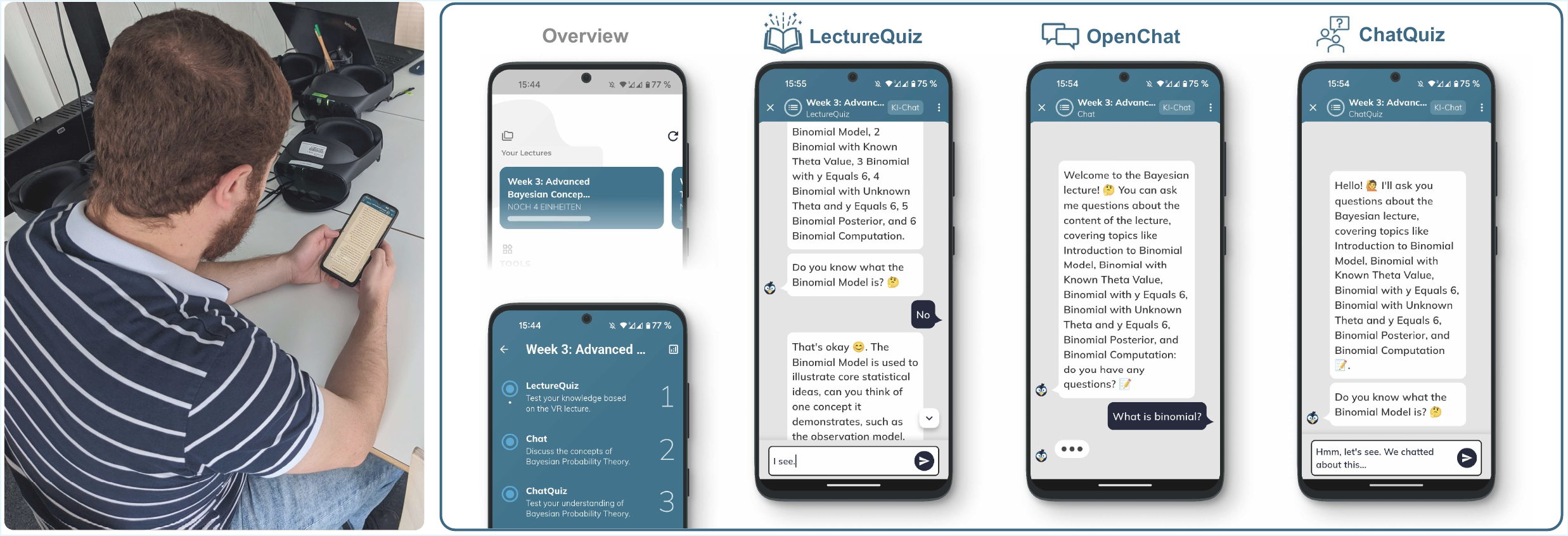}
    \caption{Integrating personalized support in the mobile assistant. Left: a student doing \textit{LectureQuiz} after a VR lecture. Right: screenshots of the assistant, including personalized \textit{LectureQuiz}, \textit{OpenChat} for Q\&A support, 
and \textit{ChatQuiz} for additional practice.}
    \Description{The figure shows a student using the mobile assistant after a VR lecture. On the left, a student is seated at a desk holding a smartphone and interacting with the LectureQuiz. On the right, four smartphone screenshots present the main features of the assistant: an overview dashboard, a LectureQuiz interface with personalized quiz questions based on lecture attention data, an OpenChat interface where students can ask questions in natural language, and a ChatQuiz interface that provides follow-up practice questions.}
    \label{fig:AMSLscreenshot}
\end{figure*}

\paragraph{LectureQuiz Integration} To situate attention-aware personalization in students’ ex-situ learning after lectures, we integrated \textit{LectureQuiz} into an existing mobile assistant application deployed for self-regulated learning at our university\new{, serving as the framework's \textit{Ex-Situ Intervention} interface.} The assistant has been implemented as a native Android and iOS application using Dart and the Flutter framework\footnote{\url{https://flutter.dev/}}. This application provides a familiar platform for students, enabling us to examine how personalized support interacts with ongoing learning practices. As described in our attention-aware personalization pipeline, the mobile application receives the JSON file with attention metrics after VR lectures. These metrics are incorporated into the prompt provided to the LLM, which then generates \textit{LectureQuiz} as a learning module within the application (see Figure \ref{fig:AMSLscreenshot}).

\paragraph{OpenChat}Beyond \textit{LectureQuiz}, the assistant integrates other supporting features, including \textit{OpenChat}, a conversational Q\&A module allowing users to pose follow-up questions and receive answers after the lecture. For \textit{OpenChat}, the mobile assistant leverages a knowledge base constructed from lecture slides, textbooks, and recorded lecture transcripts. When a student submits a question, a similarity search is performed on a PostgreSQL database using embeddings. The most relevant content is then appended to the LLM prompt as grounding information, following the retrieval-augmented generation approach.

\paragraph{ChatQuiz}Furthermore, chat logs in \textit{OpenChat} inform users' confusion levels per lecture section, following an established method of linguistic confusion detection~\cite{atapattu_identification_2019}. Later, the confusion levels can be used to generate an additional quiz module, \textit{ChatQuiz}. While \textit{ChatQuiz} explores further personalization possibilities, the main focus of our study remains on the attention-driven \textit{LectureQuiz} and the effect of our learning ecosystem \textit{AttentiveLearn} as a whole.

In our design, \textit{AttentiveLearn} bridges immersive in-lecture experiences with ubiquitous post-lecture support. By embedding attention-aware personalization into a mobile assistant, we explore beyond using attention data as one-off feedback and provide a continuous, learner-centered ecosystem for real-world learning. To ensure transparency and reproducibility, we made the system components of \textit{AttentiveLearn} open source and available as supplementary material.

\section{Field Study}
\label{sec:fieldstudy}
We evaluated \textit{AttentiveLearn} in a between-subjects field study with 36 participants (12 female, 24 male). The study aimed to investigate the impact of attention-aware post-lecture support on students' motivation, engagement, and learning outcomes in an authentic setting. The study protocol was approved by the university IRB. Due to IRB constraints, the study could not be embedded in the official curriculum or offer credit points toward students’ degrees, as this might have biased academic achievement. Instead, we recruited participants from a local university and organized a three-week lecture series on Bayesian data analysis and a final exam on the fourth week, offered as an optional non-credit course on the university’s learning platform. The lectures were based on a real-world course by \citet{vehtari_bayesian_2024}, which covers basic concepts of Bayesian data analysis including probability theory, single-parameter and hierarchical models, Bayesian inference, etc. and has been used as an open educational resource in universities.

\subsection{Study Design}
We employed a two-factor mixed design with one between-subjects factor \textit{Group} (attentive vs. non-attentive) and one within-subjects factor \textit{Week} (1–3). Participants were randomly assigned to one of the two groups. As the independent variable, the two groups differed only in the type of post-lecture support delivered through the mobile assistant: the attentive group received an attention-aware \textit{LectureQuiz}, while the non-attentive group received a non-personalized version. \new{For comparable expectations across two conditions, all participants were informed that the study involved \say{personalized learning support} and that eye-tracking data would be collected throughout the study. This provided a transparent and data privacy-compliant study design. However, the attention-based personalization mechanism for the attentive group was not revealed until the final interview. This prevented participants in the attentive group from knowing during the study that their attention data directly informed quiz generation.} We examined three sets of dependent variables:

\begin{itemize}
    \item \textbf{Engagement:} measured weekly with the short-form User Engagement Scale (UES-SF)~\cite{obrien_practical_2018}.
    \item \textbf{Motivation:} measured weekly with the Motivated Strategies for Learning Questionnaire (MSLQ)~\cite{pintrich_manual_1991}.
    \item \textbf{Learning Outcomes:} assessed with three weekly \textit{Mini-Exams} in the application and an on-site exam in week 4.
\end{itemize}

Additional behavioral data (eye-tracking in VR lectures, log data from \textit{LectureQuiz}, etc.) were collected to support and validate findings. Finally, semi-structured interviews with 10 participants provided qualitative insights.

\paragraph{Incentivization Strategy}
Because the course did not contribute credit points, we implemented an incentive scheme to approximate a typical student workload. Participants earned €15 for completing each weekly on-site lecture and the subsequent \textit{LectureQuiz}, plus €8 for completing the out-of-class activities (\textit{ChatQuiz} and \textit{Mini-Exam}) with the mobile assistant each week. Completion of the final exam yielded an additional €15. These rates were aligned with the standard hourly payment of student assistants at our university. In addition, participants could earn €2 bonuses for ranking in the top 25\% of each \textit{Mini-Exam} and the final exam. The incentive structure was intended to provide fair compensation across the four weeks, while avoiding undue pressure that might bias learning behavior.

\subsection{Participant Information}
\label{sec:partinfo}
We recruited 36 students (12 female, 24 male) from two local universities, aged 18-28 ($M = 23.61$, $SD = 3.26$). Seventeen were enrolled in a bachelor’s program, 14 in a master’s program, and five had recently completed their master’s degree. Pre-study surveys confirmed low prior knowledge of Bayesian data analysis (15 reported none, 21 little; $M = 1.42$, $SD = 0.50$, on a 5-point scale). Familiarity with VR varied: five had never used VR, 29 reported rare use (less than twice per year), and two reported occasional use (a few times per month).

\paragraph{Group Comparability}  
Participants were randomly assigned to the two groups (attentive vs. non-attentive). Balance checks on the initial sample ($n = 36$) showed no significant group differences in age ($t(34) = 0.710$, $p = 0.482$), prior knowledge ($U = 153.00$, $p = 0.753$), VR experience ($U = 153.50$, $p = 0.713$), or educational degree ($\chi^2(2) = 0.545$, $p = 0.762$). Assumptions of normality and variance homogeneity were confirmed where applicable.

Five participants dropped out after week 2 and three after week 3, leaving 28 participants (10 female, 18 male) who completed the full study. This drop-out rate was comparable to that of real lectures at our university. After dropout, group sizes remained comparable (13 attentive vs. 15 non-attentive), and the distribution of demographics and prior knowledge remained balanced.

\subsection{Task and Procedure}
\begin{figure*}[htbp]
    \centering   \includegraphics[width=\linewidth]{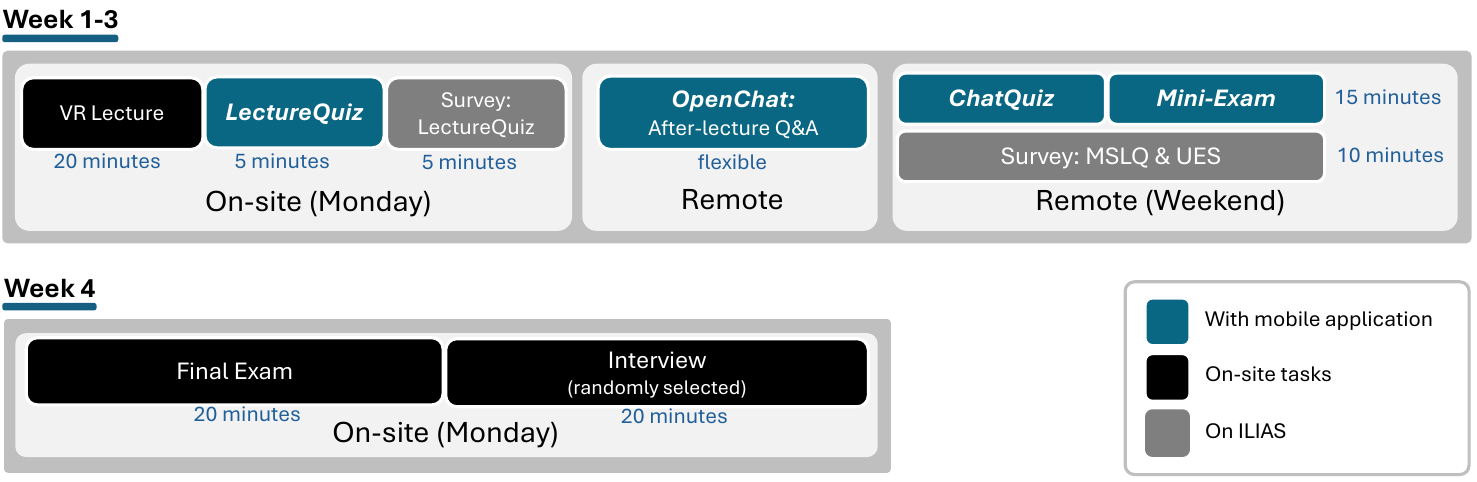}
    \caption{Study procedure over four weeks. Weeks~1–3 included a VR lecture, \textit{LectureQuiz} and survey. Followed by out-of-class \textit{OpenChat} for Q\&A, \textit{ChatQuiz}, weekly surveys, and a mini-exam. Week 4 concluded with a final exam and semi-structured interviews.}
    \Description{The figure depicts the study timeline and tasks. For Weeks 1–3, participants attended a 20-minute VR lecture on-site, followed by a 5-minute LectureQuiz and a short survey. Afterwards, they could use the OpenChat feature remotely for flexible Q\&A. During the weekend, they completed a ChatQuiz, weekly surveys including the MSLQ and UES, and a 10-minute mini-exam. In Week 4, all participants took a 20-minute final exam on-site, and a subset were randomly selected for 20-minute interviews. A legend shows that blue blocks represent mobile application tasks, black blocks indicate on-site activities, and grey blocks represent surveys administered via ILIAS.}

    \label{fig:taskfield}
\end{figure*}
Participants took part in a four-week course that combined an immersive virtual classroom with post-lecture support via the mobile assistant.

\paragraph{Weeks~1–3} Three on-site VR lectures, each around 20 minutes long, were adapted from the original Bayesian data analysis course and offered each week~\cite{vehtari_bayesian_2024}. Following the structure of the source material, each lecture was divided into six sections. The section-division was not visible to participants and had no impact on their in-lecture experience, but served as analytic units for calculating section-level attention metrics in the \textit{AttentiveLearn} backend.

Each Monday, participants attended an on-site lecture using a Meta Quest Pro headset, organized in groups of three to five to mirror the virtual classroom setting. This schedule resulted in a total of 10 on-site sessions every Monday. After the lecture, they completed a \textit{LectureQuiz} in the mobile application with six questions (personalized for the attentive group based on in-lecture attention, random for the non-attentive group). \new{Although the quizzes received by the non-attentive group were not personalized, these items were also pedagogically validated and reflect current practices in lectures: all quiz questions were generated based on pedagogically validated material including the original assessment of the open course~\cite{vehtari_bayesian_2024}, as well as the excersice questions in the course textbook~\cite{gelman_bayesian_1995}, and were reviewed retrospectively by a statistics expert and lecturer for content alignment and appropriate difficulty.} They continued with a two-item survey on their perceived accuracy and helpfulness of \textit{LectureQuiz}, measured on a 5-point Likert scale:

\begin{enumerate}
    \renewcommand{\labelenumi}{Q\arabic{enumi}.}
    \item How accurately did \textit{LectureQuiz} reflect the parts of the lecture where you paid less attention?
    \item To what extent did \textit{LectureQuiz} help you understand and review the lecture content?
\end{enumerate}

After the weekly lecture on site, participants went home and carried out all subsequent activities independently in their own environments. During the week, they could freely use \textit{OpenChat} at any time. At the end of the week, starting each Friday, a \textit{ChatQuiz} and a \textit{Mini-Exam} module (with 12 predefined questions validated by a statistics expert) were made available on the mobile assistant. Participants could complete these tasks flexibly over the weekend before the next Monday lecture. Weekly surveys on motivation (MSLQ) and engagement (UES-SF) were also administered on the online learning platform of the local university\footnote{with the built-in survey tool on the \href{https://www.ilias.de/en/}{ILIAS platform}}. This weekly cycle was repeated for three weeks (see Figure \ref{fig:taskfield}).

\paragraph{Week 4} On Monday of week 4, participants completed a final on-site exam consisting of 12 multiple-choice questions. The items were drawn from a mix of original course assessments~\cite{vehtari_bayesian_2024}, textbook exercises~\cite{gelman_bayesian_1995}, and LLM-generated questions, all of which were reviewed by a statistics expert, and none of these questions overlapped with the {Mini-Exams}. For the final exam, participants had 20 minutes and were not permitted to use any supporting materials. After the exam, and with consent, one participant per session was randomly selected for a semi-structured interview.

\subsection{Data Collection and Analysis}
\paragraph{Data Collection} 
We collected multimodal data for mixed-methods analysis: (a) eye-tracking during lectures (AOI coverage percentage, attention switches, ADI), (b) log data from the mobile assistant including \textit{LectureQuiz} scores, (c) on-site survey (two items on \textit{LectureQuiz}) and weekly surveys (MSLQ, UES-SF), (d) weekly \textit{Mini-Exams} and a final exam, and (e) post-study interviews.

For the weekly surveys, we followed established guidelines. We included the motivation scale of MSLQ with six subscales (intrinsic goal orientation, extrinsic goal orientation, task value, control beliefs, self-efficacy, test anxiety) on a 7-point Likert scale~\cite{pintrich_manual_1991}. The UES-SF measured four engagement dimensions (focused attention, perceived usability, aesthetic appeal, reward factor) on a 5-point Likert scale~\cite{obrien_practical_2018}. We also analyzed \textit{OpenChat} logs and \textit{ChatQuiz} scores from the mobile assistant to capture patterns of self-directed study.

Finally, we collected qualitative data through 10 semi-structured interviews conducted after the final exam, with five participants randomly selected from each group. The interview protocol covered four areas: (a) study routines and overall impressions, (b) perceptions of post-lecture support, (c) motivation and engagement, and (d) suggestions for system improvement. The full interview guide and coding framework are provided as supplementary materials.

\paragraph{Analysis Methods} 
Quantitative analyses combined frequentist and Bayesian approaches. Two-way mixed ANOVAs were applied for weekly surveys and exams, with \textit{Group} (attentive vs. non-attentive) as a between-subjects factor and \textit{Week} (1–3) as a within-subjects factor. Assumptions of normality, homogeneity of variances, and absence of outliers were confirmed using visual inspection of Q-Q plots and Levene's tests. \new{Furthermore, because the MSLQ and UES-SF surveys contain multiple subscales, we accounted for the risk of family-wise error by applying a Holm-Bonferroni correction strategy across the scales.} Final exam scores were analyzed using independent-samples t-tests, with assumptions of normality confirmed with a Shapiro-Wilk test ($W = 0.956, p = 0.287$) and homogeneity of variance confirmed using a Levene's test ($F(1, 26) = 1.344, p = 0.257$). \new{Additionally, to assess the construct validity of the ADI, we performed a repeated measures correlation analysis between the per-section ADI and the subjective quiz ratings on accuracy and helpfulness.}

Qualitative data from the interviews were analyzed using reflexive thematic analysis (TA). Following the six-phase approach~\cite{braun_thematic_2012}, the first author engaged in the process of familiarization with the data, initial coding, theme construction, reviewing, refining, and writing. Coding was conducted openly and inductively to remain grounded in participants’ accounts through an interpretative and iterative process, with themes generated through analytical engagement with the data. As pointed out by \citet{braun_reflecting_2019}, reflexive TA does not depend on multiple coders. Therefore, the analysis was conducted by the first author. To ensure rigor, the first author revisited codes and themes multiple times to refine their scope after distancing and engagement with the dataset. A reflexive log was kept during the process to document analytic decisions and made available as supplementary material.

\section{Results}
\label{sec:results}
We present our findings in six parts, combining quantitative and qualitative insights. First, we report on in-lecture attention patterns measured through eye-tracking in VR. Then, we analyze user feedback on the attention-aware \textit{LectureQuiz}. Next, we examine user engagement and learning motivation. Furthermore, we present learning outcomes based on weekly mini-exams and the final exam. Finally, we describe how participants appropriated the support features, including \textit{OpenChat} and \textit{ChatQuiz}, and summarize additional suggestions from the interviews.

\subsection{In-Lecture Attention}
\begin{figure*}[htbp]
    \centering
    \includegraphics[width=\linewidth]{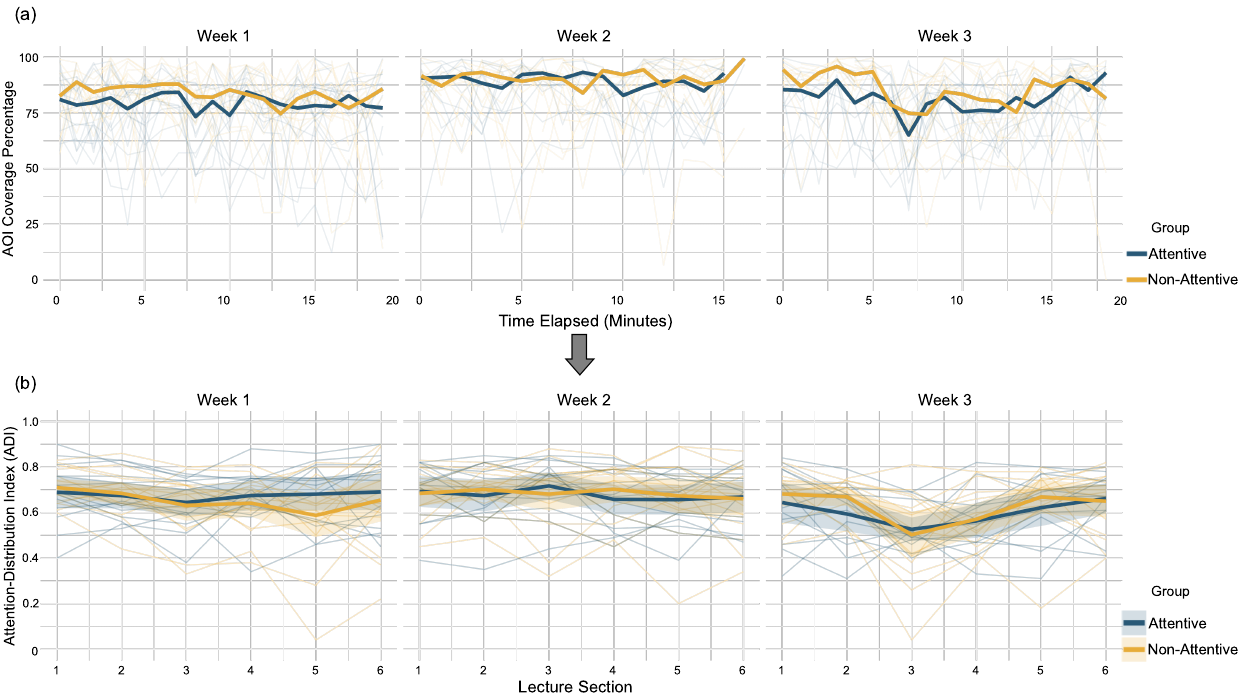}
    \caption{(a) In-lecture focused percentage over time (minutes) across three weeks for attentive and non-attentive groups; (b) Attention Distribution Index (ADI) across lecture sections for both groups.}
    \Description{Two sets of line plots compare Attentive and Non-Attentive groups over three weeks. 
    In (a), the y-axis shows focused duration percentage per minute, with individual participant lines in light colors and group averages highlighted. 
    The Attentive group (dark blue) and Non-Attentive group (gold) follow similar patterns with fluctuations across weeks. 
    In (b), the y-axis shows the Attention Distribution Index (ADI) across six lecture sections per week. 
    Lines represent participants with group averages highlighted, showing generally stable attention distribution patterns with slight differences between groups.}
    \label{fig:ETGrouped}
\end{figure*}

\paragraph{Overall Attention Distribution}
As shown in Figure \ref{fig:ETGrouped}, we visualize the eye-tracking data collected during the three-week VR lectures. The AOI coverage percentage (see Figure \ref{fig:ETGrouped}a) was calculated per minute by dividing the total gaze duration on the learning-related AOIs (lecturer avatar and lecture slides) by the 60-second duration. Our results indicate that both groups maintained similar attention levels throughout the three-week lectures.

We conducted a mixed ANOVA on the ADI metric (ranging 0--1; see Figure~\ref{fig:ETGrouped}b), calculated at the lecture-section level based on AOI coverage and attention switches. The analysis revealed a significant within-subject effect of \textit{Week}, $F(2, 52) = 5.015, p = .010, \omega^2 = 0.069$, indicating that ADI changed over time. This finding was supported by the Bayesian mixed ANOVA, which showed that the data were best explained by a model including only \textit{Week} ($BF_{10} = 9.69$), with inclusion Bayes factors providing moderate evidence for \textit{Week} ($BF_{\text{incl}} = 6.20$). Post-hoc pairwise comparisons (Holm-adjusted) localized the effect to a difference between week~2 and week~3 ($MD = 0.074, SE = 0.023; t(26) = 3.189, p_{\text{holm}} = .011; d = 0.728$), with the Bayesian comparison providing strong converging evidence ($BF_{10} = 13.08$). In contrast, both analyses provided little evidence for a main effect of \textit{Group} ($F(1, 26) = 0.177, p = .677; BF_{\text{incl}} = 0.185$) or for a \textit{Week} $\times$ \textit{Group} interaction ($F(2, 52) = 0.523, p = .596; BF_{\text{incl}} = 0.160$).

\paragraph{Participants with Low-Attention}
Recognizing variation in attention levels among participants, we conducted a post-hoc and exploratory descriptive analysis focusing on those with low attention. We defined this subset as participants whose week-1 ADI was at or below the 25th percentile ($0.619$), including eight participants evenly split across the two groups. For this subset of participants, those in the attentive group showed a greater average ADI increase ($M = +0.086$, $SD = 0.092$) than their non-attentive counterparts ($M = +0.026$, $SD = 0.187$) from week 1 to week 3. An independent-samples t-test showed that this difference was not statistically significant, $t(6) = 0.576$, $p = 0.585$. A complementary Bayesian t-test yielded $BF_{10} = 0.578$, providing anecdotal evidence for the null hypothesis and suggesting that a larger sample size would be needed to draw stronger conclusions.

\paragraph{Qualitative Results}
Thematic analysis of interview data further contextualized these findings. From 10 interviews, we identified six codes that informed \textbf{Theme 1: Promoting Attention Management in Subsequent Lectures}.

Participants described immersive VR lectures as engaging, regardless of prior VR familiarity (the interviewed participants included novice users of VR: P1, P3, P15, P24, and more frequent users: P4, P7), echoing prior work on the benefits of immersive virtual classroom~\cite{gao_digital_2021}. However, participants also highlighted challenges of attention management. Several reported initial disorientation, noting they \say{didn't know what to look for in this new type of classroom} (P24) or felt \say{lost in another world} (P3), pointing to the need for learner-centered attention support. In addition, six participants mentioned that attention was harder to sustain in week 3 as \say{the lecture became more difficult} (P1)—a pattern consistent with both course design~\cite{vehtari_bayesian_2024}, as well as the observed ADI decrease (Figure \ref{fig:ETGrouped}b). This pattern validates existing findings that attention management remains a challenge in immersive virtual classrooms~\cite{han_exploring_2022}.

While most participants agreed that \textit{AttentiveLearn} did not alter the in-lecture experience directly (since personalization occurred post-lecture), three participants in the attentive group (P7, P15, P19) reported consciously trying to sustain greater focus in later weeks, \new{motivated by anticipating the quizzes, even though the attention-based personalization mechanism was not explicitly apparent to them}. As P19 explained: \say{because I know there is a quiz afterwards and I don't want to perform bad there, I try to focus more listening (to the lecture) to make full use of the quiz}. P7 described a related effect, saying: \say{I'm aware that the quiz will help me afterwards, so I became more confident and comfortable during the lectures}. None of the non-attentive participants reported similar impressions. Importantly, these three participants from the attentive group were also among those classified as low-attentive in week 1, and their ADI scores indeed improved over time ($M = +0.13$, $SD = 0.03$).

\textbf{These results first validate existing research that attention management is a challenge in immersive learning environments. At the same time, our findings suggest that \textit{AttentiveLearn} may have motivated low-attentive students to develop more effective attention management strategies in subsequent lectures.}

\subsection{Attention-Aware \textit{LectureQuiz}}
\label{sec:quizresult}
\begin{figure}[htbp]
    \centering \includegraphics[width=\linewidth]{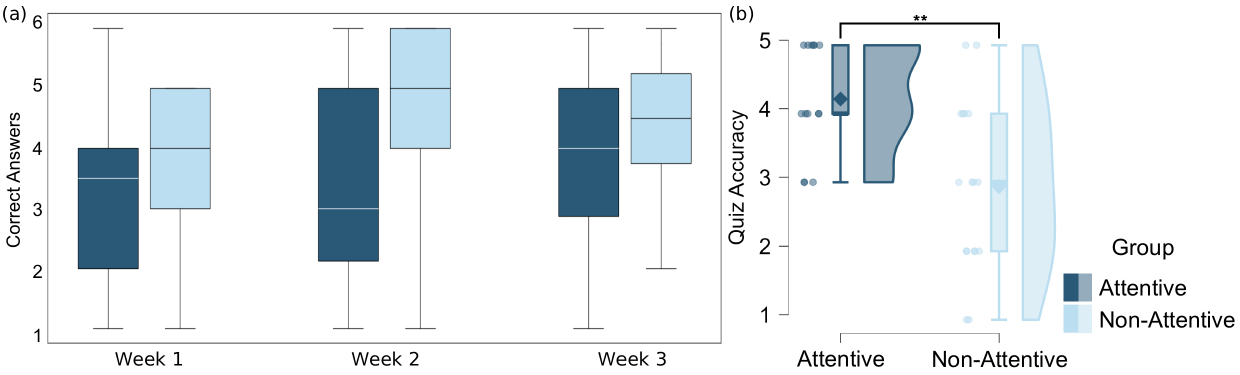}
    \caption{(a) Correct answers in \textit{LectureQuiz} across three weeks by group. (b) Aggregated \textit{LectureQuiz} accuracy ratings on a 5-point Likert scale; asterisks denote significant differences (** \textit{p} < 0.01).}
    \Description{Figure shows two panels comparing Attentive and Non-Attentive groups. Panel (a) contains three boxplots, one for each week, displaying the number of correct answers in the LectureQuiz module. Each boxplot shows the median and spread of scores for both groups. Panel (b) displays quiz accuracy ratings on a 5-point scale using a violin-boxplot. The Attentive group’s ratings cluster higher, while the Non-Attentive group’s ratings are more widely spread and lower on average. A significance bracket with two asterisks indicates a statistically significant difference between groups.}
    \label{fig:amsl:lecture_quiz}
\end{figure}

\paragraph{Quiz Scores}  
As shown in Figure~\ref{fig:amsl:lecture_quiz}, participants completed the \textit{LectureQuiz} with six questions immediately after each lecture, with responses evaluated in real time by an LLM to calculate quiz scores. Overall, the attentive group scored slightly lower across weeks ($M = 2.67$, $SD = 1.721$) compared to the non-attentive group ($M = 3.21$, $SD = 1.413$), but both frequentist and Bayesian analyses suggest that these differences were not statistically meaningful.

For the between-subject factor \textit{Group}, the mixed ANOVA indicated no significant difference, $F(1, 26) = 2.64, p = 0.11$, and the Bayesian analysis likewise provided little evidence for including \textit{Group} ($BF_{\text{incl}} = 0.396$). Similarly, for the within-subject factor \textit{Week}, no effect was observed, $F(2, 52) = 1.59, p = 0.21$, with the Bayesian analysis again indicating little support ($BF_{\text{incl}} = 0.234$). Finally, the \textit{Group} $\times$ \textit{Week} interaction was not significant, $F(2, 52) = 0.55, p = 0.58$, and the Bayesian inclusion factor strongly favored exclusion ($BF_{\text{incl}} = 0.077$).

\paragraph{Survey on \textit{LectureQuiz}}
To assess participants’ perceptions, responses from the two-item survey on quiz accuracy and helpfulness were aggregated across the three weeks. Independent-samples t-tests were then conducted.

For \textbf{quiz accuracy}, participants in the attentive group ($M = 4.21, SD = 0.80$) rated their quizzes as more accurate in revealing attention gaps than the non-attentive group ($M = 2.94, SD = 1.25$). An independent-samples $t$-test confirmed that this difference was statistically significant, $t(29) = 3.292, p = 0.003, Hedges' g = 1.157$. Consistently, the Bayesian $t$-test supported the group difference, with $BF_{10} = 14.28$. Under the directional hypothesis (attentive $>$ non-attentive), the one-sided test also supported higher perceived accuracy in the attentive group, $BF_{+0}=28.45$ (posterior median $d=1.003$). \new{Aiming to validate the objective attention metric and subjective perception, we analyzed the correlation between ADI and perceived quiz accuracy per section. For the attentive group, a repeated measures correlation indicated a significant negative association ($r_{rm} = -0.72, p < 0.001$), confirming that lecture sections with lower detected attention were significantly more likely to be rated as having accurate quizzes. In contrast, no significant correlation was found for the non-attentive group ($r_{rm} = 0.02, p = 0.714$).}

For \textbf{quiz helpfulness}, a Shapiro–Wilk test indicated non-normality ($p < 0.001$), so we used a Mann–Whitney U test, which revealed no significant difference between groups ($U = 124.000, p = 0.834$). A Bayesian Mann–Whitney U test converged, indicating that the data are about 2.8 times more likely under the null hypothesis ($BF_{10} = 0.357$). \new{Regarding helpfulness, the repeated measures analysis revealed no significant correlation with ADI for either the attentive group or the non-attentive group.}

\paragraph{Qualitative Results}
Thematic analysis provided further insights into participants’ experiences with \textit{LectureQuiz}. We generated \textbf{Theme 2: Effective Attention-Aware Personalization of \textit{LectureQuiz}} from 12 codes related to quiz accuracy, timing, and learner perceptions.

Participants across both groups valued quizzes as a form of post-lecture support. For example, non-attentive group participants noted that quizzes had helped them \say{refresh knowledge in memory} (P21). However, personalization was distinctly emphasized by attentive group participants, who described the questions as covering content that was \say{new and unfamiliar} (P7). While this sometimes led to lower quiz scores, participants interpreted it positively, recognizing that these questions reflected missed content during lectures and thus helped surface attention gaps. This aligns with the higher perceived accuracy ratings in the attentive group.

Participants further indicated a preference for attention-aware quizzes compared to potential in-lecture interventions. Several participants stated that in-lecture attention support, such as warnings~\cite{han_exploring_2022}, would have been \say{disruptive to an already complicated lecture} (P15) or introduced an undesirable \say{pressure of being monitored} (P19). In contrast, post-lecture quizzes were described as \say{just in time} (P1), making attention gaps actionable without interrupting the lecture. Finally, many participants (four out of five in the attentive group; two out of five in the non-attentive group) reported that \textit{LectureQuiz} created a positive carry-over effect with pressure, motivating them to \say{study harder through the rest of the week} (P15).

\textbf{In conclusion, attention-aware quizzes were perceived as accurate, timely, and non-disruptive, enabling participants in the attentive group to identify attention gaps and sustain learning beyond the lecture.}

\subsection{User Engagement}
\begin{figure}[htbp]
    \centering
    \includegraphics[width=\linewidth]{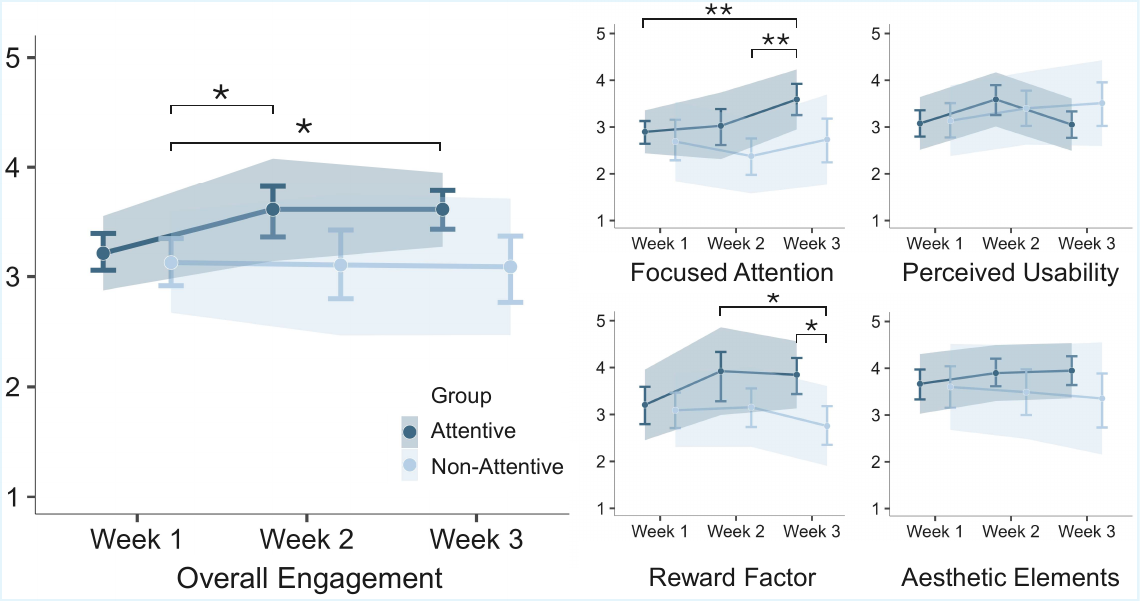}
    \caption{Engagement measured with the UES-SF scale. The plots show mean ratings for attentive and non-attentive groups across three weeks on four subscales. Shaded ribbons represent standard deviations, error bars denote 95\% confidence intervals, and brackets indicate significant group differences (* $p < 0.05$, ** $p < 0.01$).}
    \label{fig:ues}
    \Description{Figure shows five line plots comparing Attentive and Non-Attentive groups across three weeks using the UES-SF engagement scale. Panel (a) displays Overall Engagement, where the Attentive group shows higher ratings with significance brackets at Weeks 1 to 3. Panel (b) shows Focused Attention, with significant group differences at Week 2. Panel (c) presents Perceived Usability, where both groups follow similar upward trends. Panel (d) shows Reward Factor, where the Attentive group scores higher with a significant difference around Week 2. Panel (e) illustrates Aesthetic Elements, with no significant group difference. Each plot includes shaded ribbons for standard deviations, solid lines for group means, and 95\% confidence interval error bars.}
\end{figure}

\paragraph{Overall Engagement}  
To assess overall engagement, we averaged the results across the four UES-SF subscales, following \citet{obrien_practical_2018}. Descriptive analysis suggested an increase in engagement from week 1 ($M = 3.21, SD = 0.34$) to week 2 ($M = 3.61, SD = 0.47$), which was sustained in week 3 ($M = 3.60, SD = 0.32$) for the attentive group. By contrast, the non-attentive group showed a slight, non-significant decrease over the same period, from week 1 ($M = 3.13, SD = 0.46$) to week 3 ($M = 3.08, SD = 0.62$).

Before running the mixed ANOVA, Mauchly's test indicated that the assumption of sphericity was violated, $\chi^2(2) = 11.454, p = 0.003$, so a Greenhouse--Geisser correction was applied. The analysis revealed a significant between-subject effect of \textit{Group}, $F(1, 26) = 6.319, p = 0.018, \omega^2 = 0.09$, with participants in the attentive group showing higher overall engagement ($MD = 0.369, SD = 0.147$). This result was supported by the Bayesian ANOVA, which provided moderate evidence for including \textit{Group} ($BF_{\text{incl}} = 3.064$) and strong evidence for higher engagement in the attentive group ($BF_{10} = 25.94$). In contrast, neither analysis indicated meaningful effects of \textit{Week} ($F(1.462, 38.042) = 2.138, p = 0.144; BF_{\text{incl}} = 0.543$) or of the \textit{Week} $\times$ \textit{Group} interaction ($F(1.462, 38.042) = 2.909, p = 0.082; BF_{\text{incl}} = 1.061$).

\paragraph{Reward Factor and Focused Attention} 

\new{At the subscale level, analyses indicated varying degrees of evidence for between-subject effects on \textbf{focused attention} and \textbf{reward factor}.} For focused attention, the attentive group scored higher overall, $F(1, 26) = 5.768, p = 0.024, \omega^2 = 0.081, \new{p_{\text{Holm}} = 0.072}$, with Bayesian model comparison providing converging evidence ($BF_{\text{incl}} = 3.479$) and a direct comparison showing strong evidence for the attentive group ($BF_{10} = 24.65$). In addition, the within-subject effect of \textit{Week} was significant, $F(2, 52) = 5.817, p = 0.005, \omega^2 = 0.055$, supported by Bayesian model comparison ($BF_{\text{incl}} = 6.017$). Post-hoc tests showed that scores were higher in week~3 compared to week~1 ($t(26) = 3.090, p = 0.009, d = 0.483; BF_{10} = 3.361$) and compared to week~2 ($t(26) = 3.365, p = 0.007, d = 0.602; BF_{10} = 15.53$). For reward factor, the attentive group also scored higher, $F(1, 26) = 8.482, p = 0.007, \omega^2 = 0.122, \new{p_{\text{Holm}} = 0.028}$, with Bayesian model comparison providing moderate support for including \textit{Group} ($BF_{\text{incl}} = 6.622$) and a direct comparison showing strong evidence for the attentive group ($BF_{10} = 50.02$).

The other two subscales showed comparable results on perceived usability (attentive: $M = 3.329, SD = 0.607$, non-attentive: $M = 3.348, SD = 0.817$) and aesthetic elements (attentive: $M = 3.838, SD = 0.606$, non-attentive: $M = 3.482, SD = 1.026$), with no significant differences between the attentive and non-attentive groups.

\paragraph{Qualitative Results}
Our thematic analysis produced \textbf{Theme 3: Engagement Through Self-Reflection and Sustained Focus Across Weeks}. 

The codes grouped under this theme highlighted two mechanisms. First, personalized post-lecture support encouraged participants to actively reflect on their attention and learning gaps, fostering a sense of progress and reward. Second, repeated engagement with the assistant helped participants concentrate and sustain focused attention over time.

For instance, P15 described that realizing gaps after the \textit{LectureQuiz} made problems \say{visible} and encouraged him to work on them during the week. Similarly, P4 noted how daily interactions with the assistant helped him prepare for the \textit{Mini-Exam}, creating a sense of closure at the end of each week:
\begin{quote}
    After the lecture, I went home and thought about where I didn't understand. Then I would chat with it (\textit{OpenChat}) and solve some questions day by day. Over the weekend, when I did the \textit{Mini-Exam}, I felt very confident and it was satisfying to wrap up the week.
\end{quote}

Beyond this sense of fulfillment, participants also reported that the assistant helped them sustain focus during post-lecture learning and review. Participants in the attentive group not only spent more time using the assistant (see Section~\ref{sec: ChatAMSL}) but also reported that they \say{know what to ask} (P1) and felt more focused during its use.
These reflections show that participants not only perceived a reward factor but also felt more attentive with the learning assistant.

Four participants in the attentive group emphasized that this pattern became evident in week 3, when content difficulty and external exam pressure increased. For example, P19 explained that the challenges made him more \say{curious to test the capability of the learning assistant}, which helped him sustain focused attention and engagement despite pressure. By contrast, participants in the non-attentive group reported lacking this reinforcement, describing week 3 as a point where they \say{just felt like doing the bare minimum because it (the task) was intimidating} (P21).

\textbf{Therefore, post-lecture support delivered in \textit{AttentiveLearn} fostered reflection, reward, and sustained focus, enabling participants to remain engaged even under increased pressure.}

\subsection{Learning Motivation}
\begin{figure}[htbp]
    \centering
    \includegraphics[width=\linewidth]{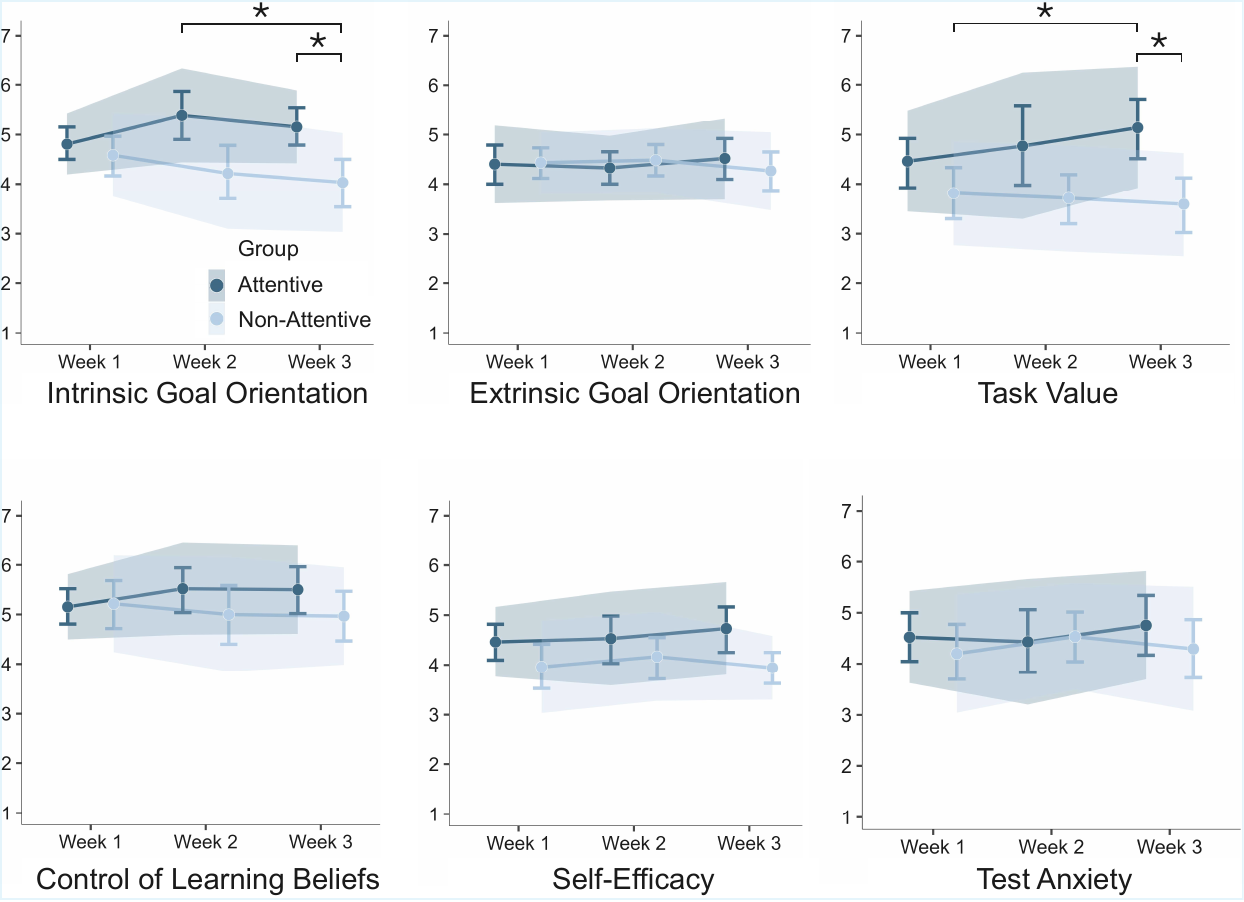}
    \caption{Mean scores on the six MSLQ subscales for attentive and non-attentive groups across three weeks. Shaded ribbons represent standard deviations, error bars denote 95\% confidence intervals, and brackets indicate significant group differences (* $p < 0.05$).}
    \Description{Figure consists of six line charts, each showing mean scores over three weeks for Attentive and Non-Attentive groups on MSLQ subscales. The y-axis ranges from 1 to 7, and the x-axis shows Weeks 1 to 3. For Intrinsic Goal Orientation, the Attentive group (solid circles) shows slightly higher scores over time compared to the Non-Attentive group (open circles). Extrinsic Goal Orientation is stable across groups. Task Value shows a visible gap, with higher scores for the Attentive group. Control of Learning Beliefs and Self-Efficacy both trend upward slightly for Attentive students. Test Anxiety remains stable across groups. Error bars represent 95\% confidence intervals.}
    \label{fig:MSLQ}
\end{figure}

To assess changes in learning motivation, a mixed ANOVA was conducted across the six MSLQ dimensions. Following \citet{pintrich_manual_1991}, we examined each dimension separately rather than calculating an overall score. Three dimensions showed significant between-group differences: \textbf{intrinsic goal orientation}, \textbf{self-efficacy}, \textbf{task value}. The assumption of sphericity was met for all analyses.

\paragraph{Intrinsic Goal Orientation}
For intrinsic goal orientation, both analyses indicated a strong between-subject effect of \textit{Group}, $F(1, 26) = 12.610, p = 0.001, \omega^2 = 0.177, \new{p_{\text{Holm}} = 0.006,}$ with the attentive group scoring higher than the non-attentive group. The Bayesian model comparison converged, providing strong evidence for including \textit{Group} ($BF_{\text{incl}} = 13.91$), and a direct group comparison showed decisive evidence for higher scores in the attentive group ($BF_{10} = 330.3$). In contrast, neither approach indicated meaningful effects of \textit{Week} ($F(2, 52) = 0.472, p = 0.626, \omega^2 = 0.000; BF_{\text{incl}} = 0.272$) or of the \textit{Week} $\times$ \textit{Group} interaction ($F(2, 52) = 3.114, p = 0.053, \omega^2 = 0.039; BF_{\text{incl}} = 0.887$).

\paragraph{Self-Efficacy} 
For self-efficacy, both analyses revealed a between-subject effect of \textit{Group}, $F(1, 26) = 5.035, p = 0.034, \omega^2 = 0.070, p_{\text{Holm}} = 0.136$, with the attentive group reporting higher self-efficacy ($M = 4.57, SD = 0.86$) than the non-attentive group ($M = 4.02, SD = 0.82$). Bayesian model comparison provided converging evidence, yielding anecdotal support for including \textit{Group} ($BF_{\text{incl}} = 1.477$), and a direct group comparison indicated strong evidence for higher self-efficacy in the attentive group ($BF_{10} = 11.23$). In contrast, neither approach indicated meaningful effects of \textit{Week} ($F(2, 52) = 0.371, p = 0.692, \omega^2 = 0.000; BF_{\text{incl}} = 0.120$) nor of the \textit{Week} $\times$ \textit{Group} interaction ($F(2, 52) = 0.752, p = 0.477, \omega^2 = 0.000; BF_{\text{incl}} = 0.096$).

\paragraph{Task Value}
For task value, there was a significant between-subject effect of \textit{Group}, $F(1, 26) = 9.586, p = 0.005, \omega^2 = 0.137, p_{\text{Holm}} = 0.025$, with higher scores in the attentive group 
($M = 4.79, SD = 1.25$) compared to the non-attentive group ($M = 3.70, SD = 1.05$). This was supported by Bayesian model comparison, which provided moderate evidence for including \textit{Group} ($BF_{\text{incl}} = 6.848$), and by a direct Bayesian comparison showing decisive evidence for higher task 
value in the attentive group ($BF_{10} = 548.12$). Neither analysis indicated effects of \textit{Week} ($F(2, 52) = 0.533, p = 0.590, \omega^2 = 0.000; BF_{\text{incl}} < 0.2$) or of the \textit{Week} $\times$ \textit{Group} interaction ($F(2, 52) = 2.074, p = 0.136, \omega^2 = 0.012; BF_{\text{incl}} < 0.2$). 

\paragraph{Qualitative Results}  
Thematic analysis further clarified how attention-aware support influenced students’ motivation, resulting in \textbf{Theme 4: Motivation Through Goal-Directed Learning and Transferable Skills}.

Participants in the attentive group emphasized that \textit{LectureQuiz} made their attention gaps explicit, which helped them define concrete goals for review. As P1 explained, identifying gaps provided \say{a clear goal of what to do after the lecture}, enabling targeted Q\&A with \textit{OpenChat}. This active goal-setting resonates with the higher self-efficacy found in survey data. Meanwhile, non-attentive participants described being \say{not sure what to do} (P3) after lectures, sometimes engaging with the assistant on less relevant or already familiar content. Their gaps often only surfaced during the \textit{Mini-Exams}, which some described as \say{frustrating} (P3).

Moreover, participants framed their motivation less around the specific lecture topic of Bayesian data analysis and more around the perceived value of refining their learning strategies. Seven of 10 interviewees stated the topic itself was not of primary interest. Yet, attentive group participants emphasized that the system still improved their motivation by helping them develop transferable skills. As P7 reflected, becoming aware of gaps and addressing them was \say{useful for real lectures and exams}, even if the content itself was not central to their studies. This interpretation aligns with survey findings of higher task value in the attentive group: students perceived the system as helping them not only retain knowledge but also improve transferrable learning skills.

\textbf{By revealing attention gaps and supporting targeted review, \textit{AttentiveLearn} improved participants' self-efficacy and perceived task value, motivating them through clearer goals and transferable learning skills.}

\subsection{Learning Outcomes}

\begin{figure}[htbp]
    \centering
    \includegraphics[width=\linewidth]{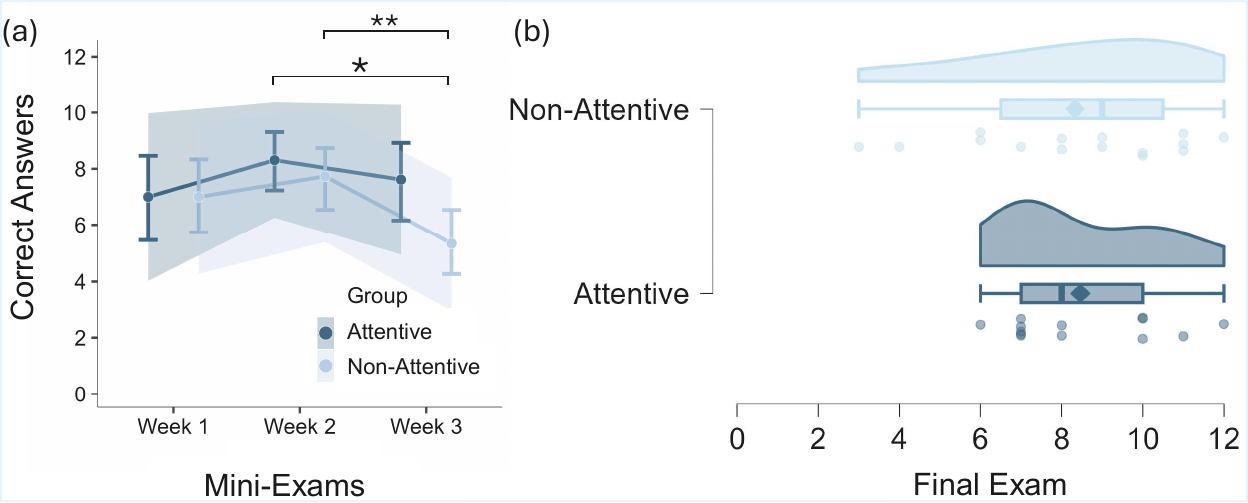}
    \caption{Performance comparison between groups. (a) Mini-exam scores across the three weeks of the study. Asterisks denote significant differences (* \textit{p} < 0.05, ** \textit{p} < 0.01). Shaded ribbons represent standard deviations; (b) Final exam scores in week~4.}
    \Description{Figure has two panels. Panel (a) is a line chart showing mean mini-exam scores across three weeks for Attentive and Non-Attentive groups. Error bars indicate variability, and asterisks mark significant differences between groups. Panel (b) is a violin-box plot showing the distribution of final exam scores in week 4 for both groups. The Attentive group is plotted below, and the Non-Attentive group above, with individual data points overlaid on the distributions.}
    \label{fig:exams}
    
\end{figure}
To examine learning outcomes, we analyzed the results of the intermediate \textit{Mini-Exams} in the mobile application and the final on-site exam.

\paragraph{Mini-Exams}  
For the mini-exams, the mixed ANOVA revealed a significant within-subject effect of \textit{Week}, $F(2, 52) = 5.989, p = 0.005, \omega^2 = 0.052$, with strong supporting evidence from the Bayesian analysis ($BF_{\text{incl}} = 11.63$). In addition, both approaches indicated a \textit{Week} $\times$ \textit{Group} interaction, $F(2, 52) = 3.413, p = 0.040, \omega^2 = 0.026; BF_{\text{incl}} = 3.81$, suggesting that performance changed differently across groups over time. Post-hoc comparisons confirmed that the non-attentive group’s scores declined significantly from week~2 ($M = 7.73, SD = 2.34$) to week~3 ($M = 5.33, SD = 2.35$), $p = 0.004$, with Bayesian tests providing strong evidence for this decline ($BF_{10} = 32.54$). In contrast, the attentive group’s scores remained stable (week~2: $M = 8.31, SD = 2.06$; week~3: $M = 7.62, SD = 2.66$). These results suggest that attention-aware personalization buffered against performance decline in later weeks.

\paragraph{Final Exam}  
For the final exam, an independent-samples t-test found no significant difference between the attentive group ($M = 8.46, SD = 1.90$) and the non-attentive group ($M = 8.33, SD = 2.69$), $t(26) = 0.143, p = 0.887$. Bayesian analysis also supported no evident difference between the groups ($BF_{10} = 0.357$), indicating the data were approximately 2.8 times more likely under the null hypothesis.

\textbf{These results show that participants benefited from attention-aware support in terms of sustaining performance during intermediate assessments, though the final on-site exam shows no significant differences.}

\subsection{Appropriation of Support Features}
\label{sec: ChatAMSL}

\paragraph{\textit{OpenChat} and \textit{ChatQuiz}}

We collected and analyzed the interaction logs of 31 participants (14 attentive, 17 non-attentive) who completed the three-week lectures and used the mobile assistant. Of these, three neither attended the final exam nor submitted the surveys, resulting in a final sample of 28 participants for the previous analyses.

On average, participants in the attentive group spent 150.83 minutes ($SD = 221.1$) per week using the mobile assistant, compared to 139.1 minutes ($SD = 340.04$) in the non-attentive group. This difference was not statistically significant. However, it may reflect higher engagement in the attentive group, consistent with the survey results.

\begin{figure*}[htbp]
    \centering
    \includegraphics[width=\linewidth]{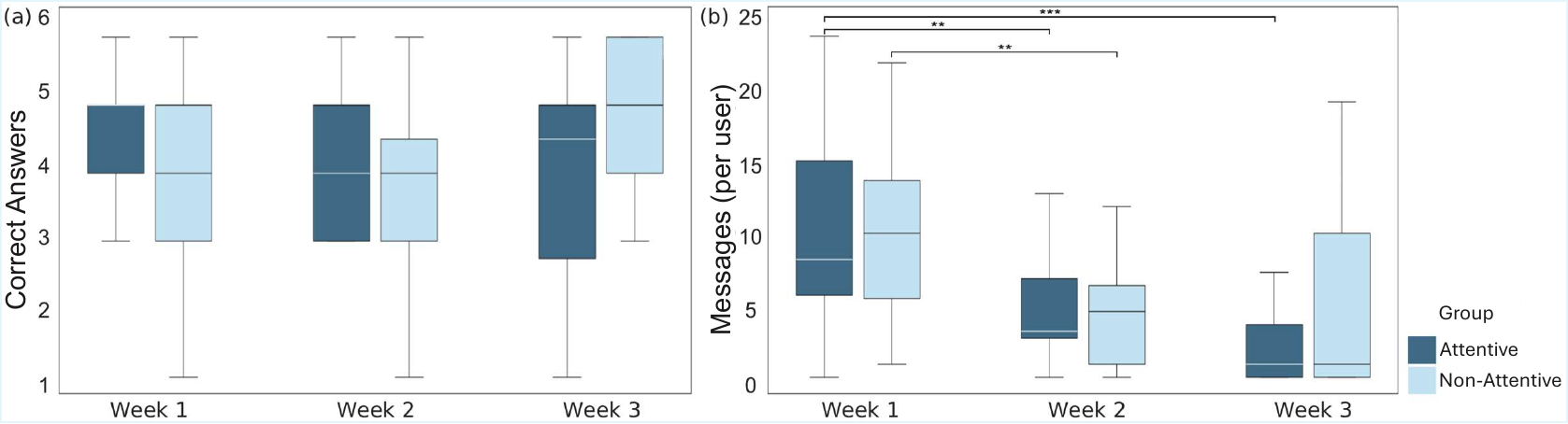}
    \caption{Boxplots of interaction data: (a) Number of correctly answered questions in \textit{ChatQuiz} module; (b) Number of messages sent in \textit{OpenChat} module per week. Asterisks denote significant differences (** \textit{p} < 0.01, *** \textit{p} < 0.001).}
    \Description{Figure shows two boxplot panels comparing groups. Panel (a) displays the number of messages sent by users per OpenChat session across groups. Panel (b) shows the number of correctly answered questions in ChatQuiz sessions across groups. Each boxplot indicates the median, interquartile range, and variability of user activity. The darker boxes represent one group, and the lighter boxes represent the other.}
    \label{fig:AMSLothers}
\end{figure*}

For \textit{OpenChat}, message activity varied significantly across weeks, $F(2, 58) = 8.91, p < .001$, but showed no group difference, $F(1, 29) = 0.32, p = .58$, and no interaction effect, $F(2, 58) = 0.26, p = .77$ (Figure~\ref{fig:AMSLothers}b). For \textit{ChatQuiz} scores (Figure \ref{fig:AMSLothers}a), mixed ANOVAs revealed no significant effects of group or week (\textit{Group}: $F(1,29) = 0.09, p = 0.77$; \textit{Week}: $F(2,58) = 0.54, p = 0.58$; \textit{Interaction}: $F(2,58) = 2.17, p = 0.13$).

\paragraph{Qualitative Results}  
Thematic analysis revealed \textbf{Theme 5: Tool Appropriation for Diverse Learning Strategies}. Five codes within this theme highlighted how participants appropriated the available features in distinct ways, reflecting their personal study practices.

For \textit{OpenChat}, some participants (e.g., P7, P15, P19) used the assistant in a highly targeted manner, entering with specific questions or hypotheses and seeking confirmation. Others—especially in the non-attentive group—reported using it more exploratively to probe for gaps and discover unfamiliar content (e.g., P3, P21). Similar appropriation patterns were observed in \textit{ChatQuiz}. Some positioned it as an exam-preparation step, completing it immediately before the weekly \textit{Mini-Exam}, while others used it diagnostically to probe their understanding before returning to \textit{OpenChat} for clarification. As P24 explained:
\begin{quote}
    I completed the (Chat)Quiz first, went back (to \textit{OpenChat}) and consolidated the knowledge. I would then take a break before the weekly exam in the app.
\end{quote}

Participants also suggested additional features to extend \textit{AttentiveLearn}, including memorization aids such as spaced repetition with flashcards (P1, P15, P21, P24), automated lecture summaries (P1, P13), and practice exam generation (P15). While beyond the scope of the current design, these suggestions reflect diverse learner strategies and highlight opportunities to expand personalization beyond attention.

\textbf{These findings suggest that \textit{AttentiveLearn} was flexibly appropriated to align with learning practices, while participants’ suggestions indicate opportunities for extending \textit{AttentiveLearn} with broader personalization.}

\section{Discussion}
\label{sec:discussion}
\subsection{Interpretation of Results}
\paragraph{Attention Improvement}

Our eye-tracking analyses confirmed that both groups began the study with comparable attention levels. The observed fluctuations in AOI coverage and ADI throughout the lecture validate existing research~\cite{han_exploring_2022}, which reveals challenges of maintaining attention in immersive VR lectures.

Although the personalized \textit{AttentiveLearn} did not produce a significant overall difference in attention trajectories between groups, exploratory analyses suggested potential benefits for low-attention participants. These students showed a tendency toward greater improvement over the three weeks when using \textit{AttentiveLearn}. The interviews provide additional context for this pattern. Participants described two possible effects: greater confidence during lectures and a stronger sense of responsibility to sustain attention in order to perform well on the quiz. Taken together, these findings tentatively indicate: \textbf{\textit{AttentiveLearn} not only facilitated post-lecture review, but may have also contributed to attention management during subsequent lectures.}

\paragraph{Motivation and Engagement}  
Quantitative measures indicate that \textit{AttentiveLearn} supported higher levels of engagement and motivation, particularly in the motivational dimensions of intrinsic goal orientation, task value, and self-efficacy, as well as the engagement factors of reward and focused attention. Other subscales did not differ across groups, which aligns with the fact that our design was not aimed to directly influence those dimensions. This demonstrates that the validated benefits of personalized quizzes, as shown in the work of \citet{mcdaniel_using_2012}, can also be extended to an attention-aware immersive learning context. \new{Because participants were not explicitly informed about the attention-based mechanism beforehand, any observed differences between groups should not be primarily attributed to participants' belief in the attentive group that they were being monitored more closely. This helps rule out expectation bias as an explanation for the observed motivational and engagement effects.}

The combined interpretation of log data, survey results, and interviews suggests that revealing attention gaps and enabling participants to address them through post-lecture support made learning challenges more visible and actionable. This process enhanced participants’ sense of task value, consistent with prior research showing that solving identifiable problems fosters motivation~\cite{pintrich_manual_1991}. In our study, this improvement in task value was achieved both through better retention of lecture content and the development of transferrable skills. 

At the same time, participants who used the personalized \textit{AttentiveLearn} reported greater self-efficacy and higher focused attention once they became aware of their attention gaps. In contrast, those in the non-attentive condition often described treating the lectures more casually, as the topic was not central to their study programs. This observation aligns with broader research on how attention-aware systems can enhance situational interest and attention management~\cite{roda_attention_2006, vertegaal_designing_2006}.

We also observed context-dependent dynamics. In week 3, coinciding with the exam phase at the universities and the increased difficulty of the VR lecture content, both groups reported higher learning pressure and lower overall MSLQ scores compared to the previous week. However, the attentive group maintained relatively higher motivation and engagement. This suggests that \textit{AttentiveLearn} helped participants regulate their motivation more effectively under heightened external demands. \textbf{Overall, \textit{AttentiveLearn} improved participants' motivation and engagement throughout the study, even when they were under pressure.}

\paragraph{Learning Outcomes}
The intermediate \textit{Mini-Exams} showed that participants using \textit{AttentiveLearn} achieved higher performance, with the performance gap widening across the three-week lectures. This effect was most pronounced in week 3, when task difficulty and external pressures were greatest. Supporting evidence comes from higher reported confidence in interviews, suggesting that personalized support improved comprehension and sustained learning progress. These findings extend prior work showing that effective quizzes enhance knowledge retention after lectures~\cite{mcdaniel_test-enhanced_2011}.

However, these benefits did not extend to the final exam. Two factors may explain this discrepancy. First, although students could revisit modules from previous weeks, \textit{AttentiveLearn} didn't provide an overarching exam-preparation feature. Without it, learners may have faced challenges to consolidate knowledge across the full course. Second, as observed in other field studies~\cite{alazmi_effects_2024, cheng_case_2019}, final learning outcomes can be heavily influenced by students' broader curriculum and other academic demands. In our study, the exam phase at the local universities constrained participants’ time for review in the final week. This challenge mirrors prior research showing that measuring long-term learning outcomes is difficult in immersive learning studies, particularly in real-world settings~\cite{petersen_pedagogical_2021}. \textbf{Therefore, \textit{AttentiveLearn} enhanced short-term comprehension and helped sustained progress, but its long-term impact was constrained and requires further investigation.}

\subsection{Limitations}
We recognize several limitations of our study. First, the sample size of 36 participants, combined with an unbalanced gender distribution, restricts the generalizability of findings. While this reflects the actual demographics of students enrolled in the two local universities, future work with a larger and more diverse sample can strengthen external validity. To mitigate the limitation, we employed Bayesian statistical analysis, a method increasingly used in immersive learning research with moderate sample sizes~\cite{liu_af-mix_2025}.

Second, the lecture length in our study was around 20 minutes. Although this aligns with existing design recommendations~\cite{meta_horizon_building_nodate}, real-world lectures often extend far longer. As current research lacks validated design guidelines for VR lecture duration, future work should investigate optimal session lengths and strategies for segmenting extended learning content for immersive learning. \new{Another pedagogical limitation is the quiz design. While all questions used in the \textit{Mini-Exams} and final exam were validated by a statistics expert and lecturer before the study, and all generated questions in the \textit{LectureQuiz} and \textit{ChatQuiz} modules were validated afterward for content alignment and pedagogical difficulty, we acknowledge that this still underrepresents the richer pedagogical intent present in real courses, where instructors hand-design non-personalized quizzes. Our comparison therefore isolates the effect of attention-aware personalization in an idealized experimental setting, rather than against a fully realistic classroom baseline.}

Third, while our learning ecosystem integrated multiple support features (e.g., \textit{OpenChat}, \textit{ChatQuiz}), our evaluation focused specifically on attention-driven personalization of \textit{LectureQuiz}. Between-subject comparisons of every design features were beyond the scope of this four-week field study. Future research should examine how different support \new{mechanisms} (e.g., attention-based and confusion-based quizzes, flashcards, tutoring, etc.) compare or complement one another when integrated into immersive learning pipelines. \new{For attention-driven personalization, we also acknowledge important limitations. While visual attention is often used as a proxy for cognitive processing under the eye–mind link hypothesis~\cite{just_theory_1980}, this assumption does not always hold. As noted in prior work \cite{szafir_artful_2013}, low attention levels indicated by biosignals (e.g., gaze, EEG) may reflect a learner's prior knowledge of the material rather than actual disengagement. To mitigate this effect, we measured participants’ prior knowledge and they reported little familiarity with the lecture content, yet residual confounds cannot be ruled out. For this, \citet{szafir_artful_2013} recommend combining multiple assessment methods and biosignals to disambiguate such cases. Therefore, we validated the attention-driven post-lecture support using eye-tracking and complemented it with quizzes as an additional assessment method. However, our current system still does not prompt users with subjective confirmations of attention gaps during learning. Future work should integrate mechanisms that couple system-driven attention signals with user-reported focus and perceived difficulties, enabling more accurate and learner-centered personalization.}

Finally, our study was conducted at our university with one lecture topic. Although this ensured ecological validity in an authentic setting, additional studies across varied domains, institutions, and cultural backgrounds are needed to establish robustness. \new{Moreover, the use of non-credit lectures and compensated participation limits the generalizability of our findings to authentic university courses where motivation, stakes, and classroom dynamics differ.}

\subsection{Design Implications for Future Research}

\paragraph{\textbf{Extending Immersive Learning with Post-Lecture Support}}
In this paper, we connected the virtual learning experience with a mobile learning assistant that delivers personalized post-lecture support. Our findings show that students in both groups continued engaging with the content after the VR session, revisiting material multiple times per week with the assistant. This indicates that even independent of gaze-adaptive personalization, extending immersive learning into post-lecture contexts increases engagement and fosters more continuous learning.

This finding aligns with ubiquitous and seamless learning frameworks~\cite{berge_seamless_2013, kloos_emerging_2011}, which emphasize that learning should not be restricted to a single context or time, but instead span across settings and devices. \new{Our findings also align with the design implications of existing research \cite{szafir_artful_2013, klaveren_effect_2017}, suggesting that revisiting content, especially when tailored to learners' attention, improves the general learning experience. Additionally, }our work demonstrates how immersive learning can be complemented by mobile assistants to bridge in-situ and ex-situ learning. As~\citet{cheng_case_2019} suggest, cross-device continuity is particularly valuable in creating an active and ubiquitous learning experience. Our findings reinforce this promise and point to design opportunities for cross-device immersive learning ecosystems.

\paragraph{\textbf{Ex-situ Personalization for Immersive Experiences}}
\textit{AttentiveLearn} implements a pipeline that processed in-situ eye-tracking data to deliver ex-situ, attention-aware personalization. This approach extends earlier visions of cross-device interaction (e.g., ~\cite{brudy_cross-device_2019}) by not only bridging modalities but also adapting user experience across devices based on users’ cognitive states. Current cross-reality research often highlights a tension between continuity and immersion, with post-experience supports potentially disrupting the sense of flow and immersion~\cite{auda_scoping_2023}. Our work demonstrates that using in-situ attention data to inform ex-situ personalization can increase engagement and motivation after the immersive session, which may address this challenge in cross-reality interaction. \new{We selected a virtual classroom as a starting point because it allowed established attention-support techniques, primarily studied in classroom or video-based learning, to be transferred into VR in a controlled and interpretable way.}

This points to a broader research agenda on cross-reality personalization strategies. Beyond gaze, additional signals could also inform the adaptation of ex-situ support features in immersive learning and related domains. Personalization does not need to be limited to attention; other cognitive and affective aspects, such as empathy~\cite{wambsganss_adaptive_2022}, may also serve as valuable input for personalized support. Prior work has demonstrated that biosignal-driven personalization can enhance user experience in diverse contexts~\cite{schultz_biosignals_2023}. Extending our pipeline to integrate such multimodal signals would enable richer personalization, offering new opportunities to investigate how ex-situ support, informed by in-situ data, fosters sustained engagement across cross-reality experiences. \new{Future work should apply this approach to more embodied and interactive immersive environments, such as virtual laboratories and makerspaces \cite{radu_virtual_2021}, skill-training scenarios with simulation \cite{zhu_designing_2025}, to understand how ex‑situ personalization can support learning beyond virtual classrooms.}

\paragraph{\textbf{Flexible Support Techniques for Different Learning Styles}}
In our study, personalized quizzes were the primary support technique, effectively revealing knowledge gaps and guiding learners to address them. While participants recognized the value of quizzes, they also expressed interest in other support techniques, such as flashcards or automatically generated example exams. These suggestions reflect heterogeneous user needs and learning preferences, consistent with prior research on individual learning strategies~\cite{makransky_immersive_2021}.

Future work can advance in two directions. First, quizzes themselves have broader potential beyond post-lecture review: prior studies show they can also enhance in-lecture engagement~\cite{raes_learning_2020}. Future immersive learning systems could therefore integrate quizzes dynamically within both in-situ and ex-situ contexts. Second, attention is only one cognitive dimension of learning~\cite{makransky_cognitive_2021}. Other aspects such as memorization, comprehension, or decision-making may be supported through complementary techniques like spaced-repetition flashcards or gamification~\cite{zhang_virtuality_2023}. Combined with our ecosystem, such techniques could create a more comprehensive support system.

\paragraph{\textbf{Improving Engagement and Motivation as Design Goals}}
While many studies in immersive learning focus on learning outcomes~\cite{radianti_systematic_2020, radu_unequal_2021,makransky_immersive_2021, baceviciute_investigating_2020, szafir_artful_2013}, our work shows the value of also targeting engagement and motivation. Indeed, learning outcomes alone may not capture the full challenges students face in immersive environments~\cite{zhang_metaverse_2022}. By designing for broader learner-centered goals such as motivation and engagement, which covers more aspects including task value, self-efficacy, and reward factors, systems such as \textit{AttentiveLearn} can improve the overall quality of the learning experience, even when the subject matter itself is not directly relevant to students’ curricula. \new{For evaluating engagement and motivation throughout the learning process, we advocate conducting field studies that assess these factors over an extended period of learning. Existing research typically investigates single-session effects and focuses primarily on recall and learning outcomes \cite{blume_students_2019, szafir_artful_2013}. In contrast, we conducted a four-week deployment within a real course context and additionally measured engagement and motivation, uncovering design implications for sustaining learning beyond the immersive session itself.}

In our \new{field} study, participants using \textit{AttentiveLearn} reported higher motivation and engagement even when they perceived the topic as peripheral to their degree programs. This suggests that designing for engagement and motivation may help sustain learner involvement across less intrinsically motivating content. Building on existing research into in-lecture support for engagement~\cite{han_exploring_2022}, future work should investigate how in-situ and ex-situ supports can be integrated to maintain motivation and engagement throughout the entire learning journey. Such integration would broaden the scope of immersive learning design, shifting from outcome-focused designs to more holistic, learner-centered experiences.

\section{Conclusion}
\label{sec:conclusion}
In this paper, we presented \textit{AttentiveLearn}, an attention-aware learning ecosystem that extends immersive learning with post-lecture support delivered through a mobile learning assistant. Leveraging attention metrics derived from eye-tracking, the system generates personalized quizzes to help learners address attention gaps and integrates these with complementary features in the mobile application. We conducted a four-week field study with university students to investigate the effects of \textit{AttentiveLearn} on motivation, engagement, and learning outcomes. Our results show that \textit{AttentiveLearn} enhanced participants’ motivation and engagement—particularly through improvements in perceived self-efficacy and task value—as well as contributed to better intermediate learning outcomes. With these findings, we contribute empirical evidence to HCI research on learner-centered support for immersive learning. While prior work has predominantly focused on in-lecture interventions, our study demonstrates the value of extending immersive learning beyond the lecture through ex-situ personalization. Looking forward, we see opportunities for personalized support with other immersive and cross-reality ecosystems. We believe this work can contribute toward making immersive learning a more ubiquitous, personalized, and learner-centered experience.

\begin{acks}
We gratefully acknowledge funding by the German Federal Ministry of Research, Technology and Space (BMFTR) for the ABBA project (Grant Number: 16DHBKI004). We also thank Jan Decker for his assistance with the system implementation.
\end{acks}

\bibliographystyle{ACM-Reference-Format}
\bibliography{references}

\end{document}